\documentclass{jfm}
\usepackage{graphicx}
\usepackage{subfigure}
\usepackage{natbib}

\ifCUPmtlplainloaded \else
  \checkfont{eurm10}
  \iffontfound
    \IfFileExists{upmath.sty}
      {\typeout{^^JFound AMS Euler Roman fonts on the system,
                   using the 'upmath' package.^^J}%
       \usepackage{upmath}}
      {\typeout{^^JFound AMS Euler Roman fonts on the system, but you
                   dont seem to have the}%
       \typeout{'upmath' package installed. JFM.cls can take advantage
                 of these fonts,^^Jif you use 'upmath' package.^^J}%
      }
  \else
  \fi
\fi

\ifCUPmtlplainloaded \else
  \checkfont{msam10}
  \iffontfound
    \IfFileExists{amssymb.sty}
      {\typeout{^^JFound AMS Symbol fonts on the system, using the
                'amssymb' package.^^J}%
       \usepackage{amssymb}%
       \let\le=\leqslant  \let\leq=\leqslant
         \let\geq=\geqslant
      }{}
  \fi
\fi

\ifCUPmtlplainloaded \else
  \IfFileExists{amsbsy.sty}
    {\typeout{^^JFound the 'amsbsy' package on the system, using it.^^J}%
     \usepackage{amsbsy}}
    {}
\fi

\newsavebox{\astrutbox}
\sbox{\astrutbox}{\rule[-5pt]{0pt}{20pt}}

\title[Constrained flow around a magnetic obstacle]{Constrained
flow around a magnetic obstacle}

\author[E. V. Votyakov, E. Zienicke and Yu. Kolesnikov]%
{
    E\ls V\ls G\ls E\ls N\ls Y\ns  V.\ns  V\ls O\ls T\ls Y\ls A\ls K\ls O\ls V$^1$
    \footnote{Current address: University of Cyprus, 75 Kallipoleos Avenue, P.O. Box 20537, 1678 Nicosia, Cyprus},\ns
      E\ls G\ls B\ls E\ls R\ls T\ns  Z\ls I\ls E\ls N\ls I\ls C\ls K\ls E$^1$\ns \break
    \and Y\ls U\ls R\ls I\ns  B.\ns   K\ls O\ls L\ls E\ls S\ls N\ls I\ls K\ls O\ls V$^2$}

\affiliation{
     $^1$Institut f\"ur Physik, Technische Universit\"at Ilmenau,
\\ PF 100565, 98684 Ilmenau, Germany\\
$^2$Fakult\"at Maschinenbau, Technische Universit\"at Ilmenau,
\\ PF 100565, 98684 Ilmenau, Germany}

    \pubyear{2007}
    \volume{##}
    \pagerange{####--####}
    \date{-- and in revised form --}

\begin{document}

\maketitle

\begin{abstract}
Many practical applications exploit an external local magnetic
field -- magnetic obstacle -- as an essential part of their
construction. Recently, \cite{Votyakov:PRL:2007} have demonstrated
that the flow of an electrically conducting fluid influenced by an
external field can show several kinds of recirculation. The
present paper reports a 3D numerical study whose some results are
compared with an experiment about such a flow in a rectangular
duct. First, we derive equations to compute analytically an
external magnetic field and verify these equations by comparing
with experimentally measured field intensity. Then, we study flow
characteristics for different magnetic field configurations. The
flow inside the magnetic gap is dependent mainly on the
interaction parameter $N$, which represents the ratio of the
Lorentz force to the inertial force. Depending on the
constrainment factor $\kappa=M_y/L_y$, where $M_y$ and $L_y$ are
half-widths of the external magnet and duct, the flow can show
different stationary recirculation patterns: two magnetic vortices
at small $\kappa$, a six-vortex ensemble at moderate $\kappa$, and
no vortices at large $\kappa$. Recirculation appears when $N$ is
higher than a critical value $N_{c,m}$. The driving force for the
recirculation is the reverse electromotive force that arises to
balance the reverse electrostatic field. The reversion of the
electrostatic field is caused by a concurrence of internal and
external vorticity correspondingly related to the internal and
external slopes in the $M$-shaped velocity profile. The critical
value of $N_{c,m}$ quickly grows as $\kappa$ increases. For the
case of well developed recirculation, the numerical reverse
velocity agrees well with that  obtained in physical experiments.
Two different magnetic systems induce the same electric field and
stagnancy region provided these systems have the same power of
recirculation given by the $N/N_{c,m}$ ratio. The 3D helical
peculiarities of the vortices are elaborated, and an analogy is
shown to exist between a helical motion inside the studied
recirculation and a secondary motion in the process of the Ekman
pumping. Finally, it is shown that a 2D model fails to properly
produce \textsl{stable} two and six-vortex structures as found in
the 3D system. Interestingly, these recirculation patterns appear
only as time dependent and \textsl{unstable} transitional states
before the Karman vortex street forms, when one suddenly applies a
retarding local magnetic field  on a constant flow.
\end{abstract}

\section{Introduction}

An electrically conducting fluid flow influenced by a local
external magnetic field is of considerable fundamental and
practical interest. Applied to the flow, a transverse homogeneous
magnetic field creates a so-called magnetic obstacle, i.e. a
region where the flow motion is retarded by the Lorentz force.

\begin{figure}
    \centerline{\includegraphics[width=10cm, angle=0,
    clip=yes]{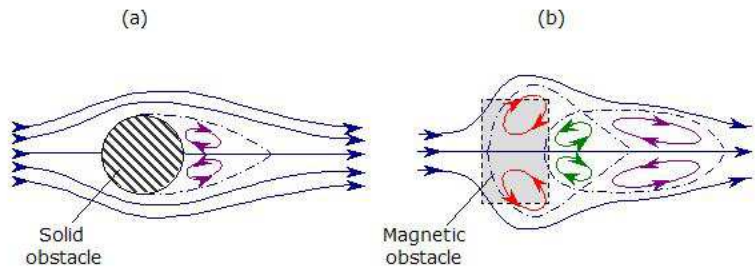}}
    \caption{\label{Fig:Introductory}Structure of the wake of a solid ($a$)
    and magnetic obstacle ($b$). By forming the wake, the solid obstacle
    develops attached vortices, while the magnetic obstacle develops inner (first pair),
    connecting (second) and attached vortices (third pair).}
\end{figure}

On the fundamental side, such a system possesses a rich variety of
dynamical states. This can already be imagined  from the fact that
its behavior is characterized by two parameters, the Reynolds
number $Re=u_0H/\nu$ and interaction parameter $N=\sigma
HB_0^2/\rho u_0$, where $H$, $u_0$, $B_0$ are characteristic
scale, velocity and magnetic field induction, and $\rho$, $\nu$,
$\sigma$ are density, kinematic viscosity, and electric
conductivity of the fluid, see e.g.
\cite{Shercliff:book:1962,Roberts:1967,Moreau:book:1990,
Davidson:book:2001}. $Re$ represents the ratio of the inertial to
viscous forces in the flow, and $N$ represents the ratio of the
Lorentz forces to the inertial forces. In ordinary hydrodynamics,
such as a flow around a solid obstacle, an increase of the
inertial force, i.e. $Re$, renders a system to the nonlinear
dynamics characterized by a vortex motion past the obstacle, see
Fig.~\ref{Fig:Introductory}$a$. The additional flow parameter $N$
brings new nonlinear degrees of freedom into the problem as was
elaborated recently by \cite{Votyakov:PRL:2007}, see
Fig.~\ref{Fig:Introductory}$b$.

On the practical side, spatially localized magnetic fields play an
essential role in a variety of industrial applications in
metallurgy, e.g. \cite{Davidson:Review:1999}, including stirring
of melts by a moving magnetic obstacle (called electromagnetic
stirring, e.g. \cite{StirringKunstreich:2003}), removing undesired
turbulent fluctuations during steel casting using steady magnetic
obstacles (called electromagnetic brake, e.g.
\cite{Takeuchi:Proc:2003}) and non-contact flow measurement using
a magnetic obstacle (called Lorentz force velocimetry, e.g.
\cite{Thess:Votyakov:Kolesnikov:2006}). For instance, it is
important to understand whether the useful turbulence-damping
effect of a magnetic brake is obliterated by excessive vorticity
generation in the wake of the magnetic obstacle.

As is well known in the flow past a solid obstacle, there is a
stagnancy region where one can observe a recirculation in the
appropriate Reynolds number regime, so called attached vortices
shown in Fig.~\ref{Fig:Introductory}$a$. If instead of the solid
obstacle one places a magnetic obstacle, by means of a local
external magnetic field, then there will appear electrical eddy
currents $\bf{j}$ which induce the Lorentz force $\bf{F_L=j\times
B}$. The largest retarding effect occurs where the transverse
magnetic field ${\bf B}$ is a maximum. Therefore, past the
magnetic obstacle there is also a stagnancy region where a kind of
a reverse flow might be obtained. This analogy between a solid and
magnetic obstacle has been realized from the beginning of
magnetohydrodynamics (MHD). In the earliest 2D numerical
simulation, \cite{Gelfgat:Peterson:Sherbinin:1978} have observed a
kind of recirculation and marked an analogy with a solid body:
\textit{'the qualitative pattern of the streamlines in such a flow
is similar to the situation which arises in the case of the flow
around objects'}. However, the specially designed physical
experiments of \cite{Gelfgat:Olshanskii:1978} following upon this
simulation failed to confirm their previous numerical results:
\textit{'special attempts which we made to detect zones with
return flow were not successful. The negative flows which occur in
certain numerical calculations are obviously  due to inaccuracies
in the calculation'}. The authors of this quotation have been
correct about the inaccuracies of the refereed 2D approach in the
sense that this approach is not suitable to describe their
experiments, nevertheless there have been still chances to reveal
a kind of recirculation in the experiments. We shall discuss this
item later in the Section \ref{sec:centerlines}.

For Western readers, the term `magnetic obstacle` has been revived
in the work \cite{Cuevas:Smolentsev:Abdou:2006}. (In the seventies
of last century, one of the authors, Yu.K., used `magnetic
obstacle` as a working term in Riga, MHD center of the former
USSR). \cite{Cuevas:Smolentsev:Abdou:2006} have performed a 2D
numerical study and  described a Karman vortex street past the
obstacle similar to those observed past a circular solid cylinder.
We discuss a link between our 3D stationary and their 2D
nonstationary results in Section \ref{sec:3dversus2d}.

Most recent results about the wake of a magnetic obstacle have
been obtained by \cite{Votyakov:PRL:2007}. Their main result is
presented in Fig.~\ref{Fig:Introductory} where the qualitative
structure of the wake of a magnetic obstacle is given in
comparison with the wake of a physical  obstacle. By means of 3D
simulation and experiments, \cite{Votyakov:PRL:2007} have found
that the liquid metal flow shows three different regimes: (1) no
vortices, when viscous force prevails at small Lorentz force, (2)
one pair of \textit{inner magnetic} vortices between the magnetic
poles, when Lorentz force is high and inertia small, and (3) three
pairs, namely, magnetic as in (2), \textit{connecting} and
\textit{attached} vortices, when Lorentz and inertial forces are
high. The latter six-vortex ensemble is shown in
Fig.~\ref{Fig:Introductory}$b$.

An important factor for the flow influenced by an external
magnetic field is the spanwise heterogeneity of the field. One can
distinguish  two extremal cases: (i) pointwise braking Lorentz
force, and (ii) spanwise homogeneous braking Lorentz force. The
latter case can be easily created, e.g. by external magnets long
enough to overlap the duct, while the first case represents an
idealization since it is impossible to have a pointwise external
magnetic field. The first case is well studied in a 2D stratified
flow, e.g. \cite{Voropayev:Afanasyev:Book:1994}; it is shown that
a dipolar vorticity is generated in the vicinity of the origin of
the point braking force. Applied to MHD flows  similar results
were obtained with a 2D numerical simulation of a creeping flow
both in \cite{Gelfgat:Peterson:Sherbinin:1978}  and recently by
\cite{Cuevas:Smolentsev:Abdou:PRE:2006}. The vortex dipole
observed in those works is of the same nature as the magnetic
vortices that we will discuss in the present paper. The second
case, spanwise homogeneous magnetic fields, is of traditional MHD
interests and  well understood. In particular, $M$-shaped profile
is developed under streamwise magnetic field gradient,
\cite{Shercliff:book:1962}. It was studied extensively in
experiments (\cite{Kit:Peterson:Platnieks:1970},
\cite{Tananaev:book:1979}) and by numerical simulation
(\cite{Sterl:1990},\cite{Votyakov:Zienicke:FDMP:2006}). The most
recent numerical paper (\cite{Votyakov:Zienicke:FDMP:2006}) by
comparing with experiment (\cite{Andrejew:Kolesnikov:2006})  has
established that when turbulent pulsations are suppressed by an
external magnetic field, it is the interaction parameter $N$ which
governs the flow. It has been  shown numerically that a spanwise
homogeneous magnetic field is not able to reverse electric field
inside the magnetic gap. As we shall prove below this is a
necessary condition to induce recirculation  between magnetic
poles.

The goals of the present paper are the following: (1) report
details not published in \cite{Votyakov:PRL:2007} due to lack of
space, (2) investigate thoroughly how a constrainment of MHD flow
influences stationary vortex patterns, (3) explain the driving
force for the recirculation,  (4) find out what are the 3D
peculiarities of the flow, and (5) clarify whether a 2D flow
contains the observed stationary vortex patterns. It will be
demonstrated that the decisive parameter for the constrained MHD
flow is a power of the recirculation between magnetic poles given
by the $N/N_{c,m}$ ratio, where $N_{c,m}$ is a critical value of
the interaction parameter to induce magnetic vortices at the given
magnetic field configuration. Moreover, several successful
comparisons with available experimental data will be given for the
intensity of the magnetic field, and for a maximal stationary
reverse flow inside the magnetic obstacle. It will be explained
that the magnetic vortices firstly appear due to the reverse
electromotive force which is induced in order to balance the
reverse electrostatic field inside the magnetic obstacle. Also, we
will discuss a 3D versus 2D numerical approach and show that the
found vortices have a 3D helical structure, while in a 2D model,
at the given range of parameters, these vortices are not fixed by
magnetic field and generate vortex shedding.

The subject of the present paper has a close connection to
questions of stability of MHD flows, however, we have not included
a full bifurcation analysis of new stationary flow patterns. The
 paper sheds light on the physical factors that determine
the occurrence of stationary recirculation, i.e. the spanwise
inhomogeneity of the magnetic field and the necessity of a
three-dimensional geometry. We consider the bifurcation and
stability analysis as a further step, which
--- from the practical point of view --- also involves additional
programming work on our code to allow for the computation of
Jacobi matrices and their eigenvalues in a high dimensional
dynamical system. Therefore, besides others, especially the
following question will remain open: are the topological changes
of the flow patterns, which we have observed in changing the
system parameters, caused by changes of stability, i.e.
bifurcations, or not (i.e. staying on the same solution branch but
the solution only changes topology)? This and other open questions
surely deserve further investigation for MHD channel flow in
inhomogeneous magnetic field.

The structure of the present paper is as follows. Section
\ref{sec:problem} presents a sketch of the model, equations and a
3D numerical solver. As an essential part, it describes in Section
\ref{sec:mf} an analytical method to deal with magnetic field of
arbitrary configuration.  Section \ref{sec:results} presents the
results of our numerical simulations: stationary flow patterns in
the middle plane in Section~\ref{sec:2dresults}, stability diagram
in Section \ref{sec:diagram}, mechanism for recirculation in
Section \ref{sec:vorticity}, the 3D peculiarities of vortices in
Section \ref{sec:3dresults}, as well as a relationship between 3D
and 2D numerical methods in Section \ref{sec:3dversus2d}. Finally,
the paper ends with a conclusion of the observations.

\section{Problem Definition}\label{sec:problem}

%\section{Model, equations, 3D numerical method}\label{sec:eq}

\subsection{Model, equations, numerical method}\label{sec:eq}

\begin{figure}
    \centerline{\includegraphics[width=10cm, angle=0,
    clip=yes]{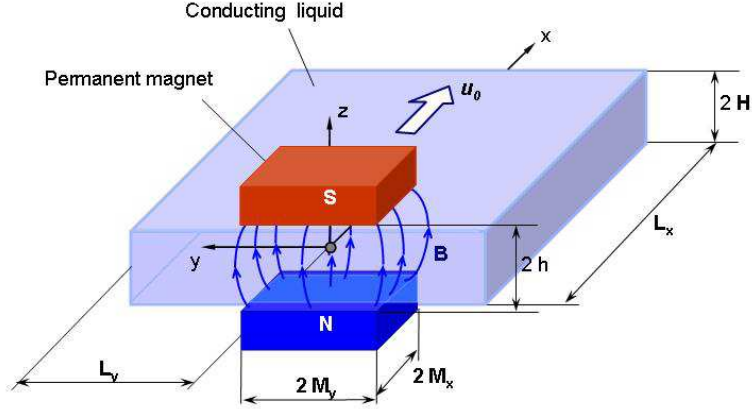}}
    \caption{\label{Fig:Sketch}Sketch of the model. Through the
    paper it is used the constrainment factor $\kappa=M_y/Ly$, where $M_y$
    and $L_y$ are half-width of the magnet and duct, correspondingly.}
\end{figure}

A schematic of the model is presented in Fig.~\ref{Fig:Sketch}: a
conducting fluid flows in the rectangular duct of dimensions
$\mbox{Length}\times\mbox{Width}\times\mbox{Height}=2L_x\times
2L_y\times 2H$ (half-length of the duct is shown); $x$-axis
corresponds to the main direction of the flow. Top, bottom and
side walls of the duct are no slip and electrically insulating.
The magnets of horizontal dimensions
$\mbox{Width}\times\mbox{Length}=2M_y\times 2M_x$ are assembled
symmetrically on the top and bottom walls, where $2h$ is the
distance between north and south poles. The center of the magnetic
gap is the center of the coordinate system. The constrainment
factor $\kappa=M_y/L_y$ defines the spanwise distribution of the
magnetic field; $\kappa$ is the varied geometric parameter in the
present simulations. Below, we will refer to the case of
$\kappa=0.02$ as a magnetic blade, $\kappa=0.4$ as a middle
magnet, and $\kappa=1.0$ as a broad magnet.

If nothing else is specified, throughout the paper the following
geometric parameters have been taken: $L_x\!=\!25$,$\;L_y\!=\!5$,
$\;H\!=\!1$, $\;h\!=\!1.5$, $\;M_x=1.5$,
$\;0.02\!\le\!\kappa\!\le\!1$. Reynolds number, $Re$, and
interaction parameter, $N$, are defined with half-height of the
duct $H$, the mean flow rate $u_0$, and magnetic field intensity
$B_0$ taken at the center of the magnetic gap,
$x\!=\!y\!=\!z\!=\!0$.

The governing equations for electrically conducting and
incompressible fluid are derived from the Navier-Stokes equation
coupled with the Maxwell equations for moving medium and  Ohm's
law. We apply the quasi-static (induction-less) approximation
where it is assumed that an induced magnetic field is infinitely
small in comparison to the external magnetic field, see, e.g.
\cite{Roberts:1967}, and is therefore neglected when one
calculates the Lorentz force, but it is not neglected when finding
the electric current density {\bf j}. The resulting equations in
dimensionless form are:
    \begin{eqnarray}
    \label{eq:NSE:momentum}
    \frac{{\partial\textbf{u}}}{{\partial t}} + (\textbf{u} \cdot \nabla ) \textbf{u}
    &=& - \nabla p + \frac{1}{Re}\triangle \textbf{u} + N
    (\textbf{j}\times\textbf{B}),
%    \\
    \label{eq:NSE:continuity} \;\;\; \nabla \cdot \textbf{u} = 0,
    \\
    \label{eq:NSE:Ohm} \textbf{j} &=&   -\nabla\phi + \textbf{u} \times
    \textbf{B},
    \label{eq:NSE:Poisson} \;\;\;\; \nabla \cdot \textbf{j} = 0,
%   \\
%    \label{eq:NSE:Poisson} \triangle \phi = \nabla \cdot(\textbf{u}\times\textbf{B}).
    \end{eqnarray}where $\bf{u}$ denotes velocity field, $\bf{B}$ is an external
magnetic field, $\bf{j}$ is electric current density, $p$ is
pressure, $\phi$ is electric potential. The  interaction parameter
$N$ and Reynolds number $Re$, $N=Ha^2/Re$, are linked by means of
the Hartmann number $Ha$: $Ha=HB_0(\sigma/\rho\nu)^{1/2}$. The
Hartmann number determines the thickness of Hartmann boundary
layers, $\delta/H \sim Ha^{-1}$ for the flow under constant
magnetic field. For a given conducting fluid and geometry of the
duct, one varies either the flow rate velocity  $u_0$, i.e. $Re$,
or the magnetic field intensity $B_0$, i.e. $Ha$. In both cases,
$N$ changes.

At given external field $\bf{B}$, the unknowns of the partial
differential equations (\ref{eq:NSE:momentum} --
\ref{eq:NSE:Poisson}) are the velocity vector field ${\bf
u}(x,y,z)$, and two scalar fields: pressure $p(x,y,z)$ and
electric potential $\phi(x,y,z)$. For no-slip and insulating
walls, the boundary conditions are $u|_\Gamma=0$,
$\partial\phi/\partial{\bf n}|_\Gamma=0$, where ${\bf n}$ is
normal vector to a surface $\Gamma$. The outlet of the duct was
treated as a force free (straight-out) border for the velocity.
The electric potential at inlet and outlet boundaries was taken to
be equal to zero because the inlet and outlet are sufficiently far
from the region of magnetic field. The stationary laminar profile
of an infinite rectangular duct known analytically in the form of
a series expansion was used as the inlet velocity  profile.
Because we are interested in a stationary solution, the initial
conditions play no role (except for the speed of convergence).

The 3D numerical solver has been described in details in
\cite{Votyakov:Zienicke:FDMP:2006}. It was developed from a free
hydrodynamic solver
%Nast3DGP\footnote{http://wissrech.iam.uni-bonn.de/research/projects/NaSt3DGP/index.htm}
created originally in the research group of Prof.~Dr.~M.~Griebel
(\cite{Griebel:book:1995}). The solver employs the Chorin-type
projection algorithm and finite differences on an inhomogeneous
staggered regular grid. Time integration is done by the explicit
Adams-Bashforth method that has second order accuracy. Convective
and diffusive terms are implemented by means of the VONOS
(variable-order non-oscillatory scheme) scheme. The 3D Poisson
equations for pressure and electric potential, arising at each
time step, are solved by using the bi-conjugate gradient
stabilized method (BiCGStab).

The computational domain, $|x|\!\leq\!L_x$, $\;|y|\!\leq\!L_y$,
$\;|z|\!\leq\!H$, has been discretized by an inhomogeneous regular
3D grid  described in detail in
\cite{Votyakov:Zienicke:FDMP:2006}. To verify that the inlet and
outlet boundaries have no influence on the presented results, we
have carried out several simulations with double the number of
grid points in the $x$-direction and  found no differences.
Moreover, we have also varied  the inhomogeneous grid resolution
both in $y$ and $z$-direction to be sure that Hartmann and
sidewall layers are properly resolved.

\subsection{Fast analytical method for a proper magnetic field}\label{sec:mf}

As is imposed by the electrodynamics, an external magnetic field
must be divergence and curl-free. Although authors of previous
works realized this fact, to define their fields they used simple
mathematical functions which did not satisfy divergence and/or
curl-free requirements, see, e.g. \cite{Sterl:1990},
\cite{Alboussiere:2004}. This fact might be explained by many
reasons, e.g. either by the complexity of a real field or
insignificance of effects appearing due to an inaccuracy of the
field definition. Therefore  we feel that there are insufficient
correct and yet simple methods to define an external magnetic
field of arbitrary configurations.  To fill this gap we explain
below a simple physical approach which can be easily extended and
implemented into 3D numerical models.

We assume that a magnet is composed of perfectly aligned pointwise
magnetic dipoles having the same magnetic moment. This assumption
is well posed for modern manufactured permanent magnets, as
follows from the final comparison between calculated and
experimentally measured magnetic field. Take $z$ as a direction of
the unit magnetic dipole ${\bf m}=(0,0,1)$, then, a partial
magnetic field in point ${\bf r}=(x,y,z)$ from a dipole located in
point ${\bf r'}=(x',y',z')$ can be presented as, see
e.g.~\cite{Jackson:Book:1999}:
\begin{eqnarray}\label{eq:MF:dipole}
    \bf{B'(r,r')}
    &=& \nabla\times\left[{\bf m}\times\frac{{\bf R}}{R^3}\right]
    =  \nabla \times\left[ - {\bf m} \times \left(
    \nabla\frac{1}{R}\right) \right] \nonumber \\
    &=& \nabla\times\left[
    \nabla\times\left( \frac{{\bf m}}{R}\right) -
    \frac{1}{R}\nabla\times{\bf m}
    \right]
    = \nabla\left[ \nabla\cdot\frac{{\bf m}}{R}\right]
    = \nabla \frac{\partial}{\partial z}\frac{1}{|{\bf r-r'}|},
\end{eqnarray} where ${\bf R}={\bf r-r'}$ and $R=|{\bf r-r'}|$. Here we used  ${\bf R}/R^3=
-\nabla(1/R)$ along with few vector identities and omitted the
constant $\mu_0/(4\pi)$. Then, the total field from a magnet
occupying a space $\Omega$ follows as:
\begin{eqnarray} \label{eq:MF:total}
    {\bf B(r)}\!=\!\int_\Omega {\bf B'(r,r')}d{\bf r'}
    \!=\! \nabla \frac{\partial }{\partial z}\Phi({\bf r}), \;\;
    \Phi({\bf r})=\int_\Omega \frac{d {\bf r'}}{|{\bf r-r'}|}
\end{eqnarray}

The last integral can be computed analytically in some cases as we
show below. For an arbitrary $\Omega$, the integration can be
performed only numerically and then be once tabulated in a 3D
array. This 3D array can be supplied into a numerical solver where
a finite differentiation is applied to compute the external
magnetic field. Another way is to differentiate analytically
$1/|{\bf r-r'}|$ and then calculate numerically three integrals
for each magnetic field component.

When limits of the integration imposed by $\Omega$ are independent
of each other, then the problem has an analytical solution by
means of the indefinite integrals given in
Appendix~\ref{ap:intergrals}.

As shown in Fig.~\ref{Fig:Sketch}, the magnetic dipoles are
located in the region $\Omega=\{|x'|\leq M_x, |y'|\leq M_y, h \leq
|z'| \leq \infty\}$, where $2h$ is a distance between north and
south magnetic poles. In the present paper, the condition
$|z'|\geq h$ is assumed because the magnets used in the
experiments are assembled onto a soft-iron yoke that closes
magnetic field lines, i.e. the dipoles are effectively located
from $h$ up to infinity. By taking the corresponding derivatives
of indefinite integrals given in Appendix~\ref{ap:intergrals}, and
after a few algebraic calculations one obtains:
\begin{eqnarray} \label{eq:MF:final}
    B_{x}(x,y,z)&=& \frac{1}{B_{0}} \sum_{k=\pm 1} \sum_{j=\pm 1}\sum_{i=\pm 1}
    (ijk)\,\mbox{arctanh}\left[\frac{y-jM_y}{r(i,j,k)}\right], \label{eq:MF:Bx}\\
    B_{y}(x,y,z)&=& \frac{1}{B_{0}} \sum_{k=\pm 1} \sum_{j=\pm 1}\sum_{i=\pm 1}
    (ijk)\,\mbox{arctanh}\left[\frac{x-iM_x}{r(i,j,k)}\right], \label{eq:MF:By}\\
    B_{z}(x,y,z)&=& -\frac{1}{B_{0}} \sum_{k=\pm 1} \sum_{j=\pm 1}\sum_{i=\pm 1}
    (ijk)\,\mbox{arctan}\left[\frac{(x-iM_x)(y-jM_y)}{(z-kh)r(i,j,k)}\right], \label{eq:MF:Bz}
\end{eqnarray} where
$r(i,j,k)=[(x-iM_x)^2+(y-jM_y)^2+(z-kh)^2]^{1/2}$, and $B_{0}$ is
selected in such a way to have $B_z(0,0,0)=1$. Three-fold
summation with the sign-alternating factor $(ijk)$ reflects the
fact that these equations are obtained by integrating according to
Eq.~(\ref{eq:MF:total}).

\begin{figure}
    \centerline{\includegraphics[width=13.75cm, angle=0,
    clip=yes]{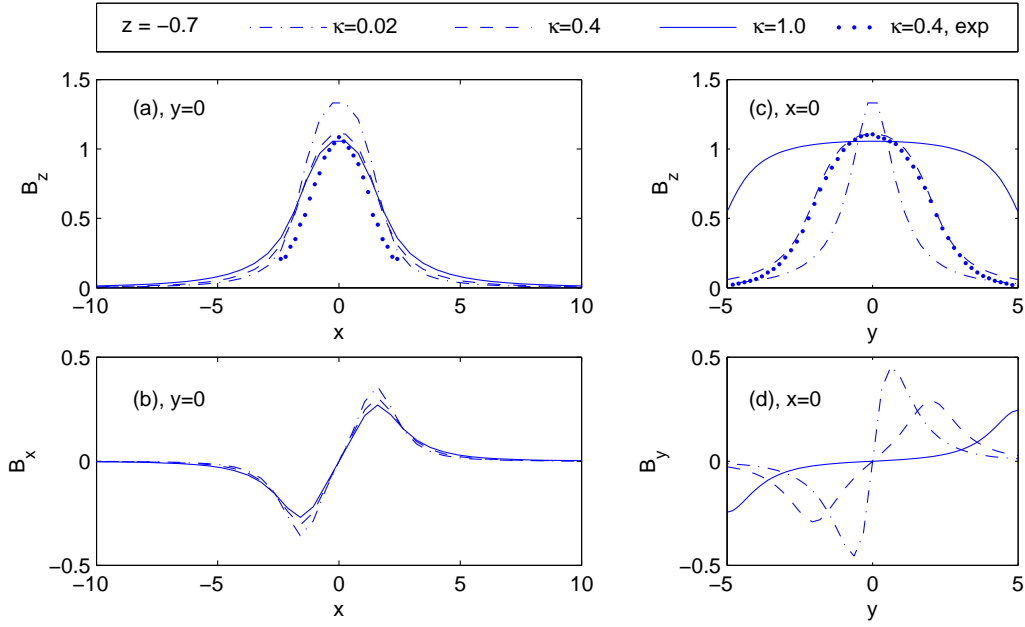}}
    \caption{\label{Fig:MF}Comparison of computed (lines) and experimental (symbols)
    magnetic field intensities at different  $\kappa=M_y/L_y$, $\kappa=0.02$
    (dot-dashed), $0.4$ (dashed and symbols) and 1.0 (solid), $z=-0.7$:
    $B_z$ component along $x,\;y=0$, ($a$),
    $B_x$ component along $x,\;y=0$, ($b$),
    $B_z$ component along $y,\;x=0$, ($c$),
    $B_y$ component along $y,\;x=0$, ($d$).}
\end{figure}

Fig.~\ref{Fig:MF} shows cuts of magnetic field intensities
computed with equations (\ref{eq:MF:Bx}-\ref{eq:MF:Bz}) for
different $\kappa$. Also, few experimental data (symbols) are
presented for $\kappa=0.4$. There is a good agreement between
experimental and analytical results. Moreover, one can see that
the constrainment factor $\kappa$ affects mainly the spanwise
distribution of magnetic field, whereas streamwise distribution
changes only slightly. This is expected, because the length $M_x$
of the magnet is fixed, while the width $M_y=\kappa L_y$ is
varied. It is necessary to note that even for the broad magnet
($\kappa=1.0$) there is still a decline of $B_z$ near side walls,
Fig.~\ref{Fig:MF}($c$).

Contour lines of the $B_z$ component in the middle plane ($z=0$)
at different $\kappa$ are also shown  in
Fig.~\ref{Fig:Streamlines}($a-c$).

Thus, we have demonstrated a self-consistent analytic approach to
define arbitrary  magnetic field configurations. It guarantees
divergence- and curl-free requirements of ${\bf B(r)}$ and has a
link with a clear physical model.

\section{Results and Discussion} \label{sec:results}

\subsection{Stationary flow patterns in the middle plane} \label{sec:2dresults}

In  Section \ref{sec:2dresults} we discuss characteristic
stationary flow patterns which have been extracted from 3D
numerical results. Three-dimensionality of the simulations is of
importance to make these patterns stable, as we shall show later
in Section \ref{sec:3dversus2d}.

\subsubsection{Streamlines for different constrainment $\kappa$}

\begin{figure}
    \centerline{\includegraphics[width=13.75cm, angle=0,
    clip=yes]{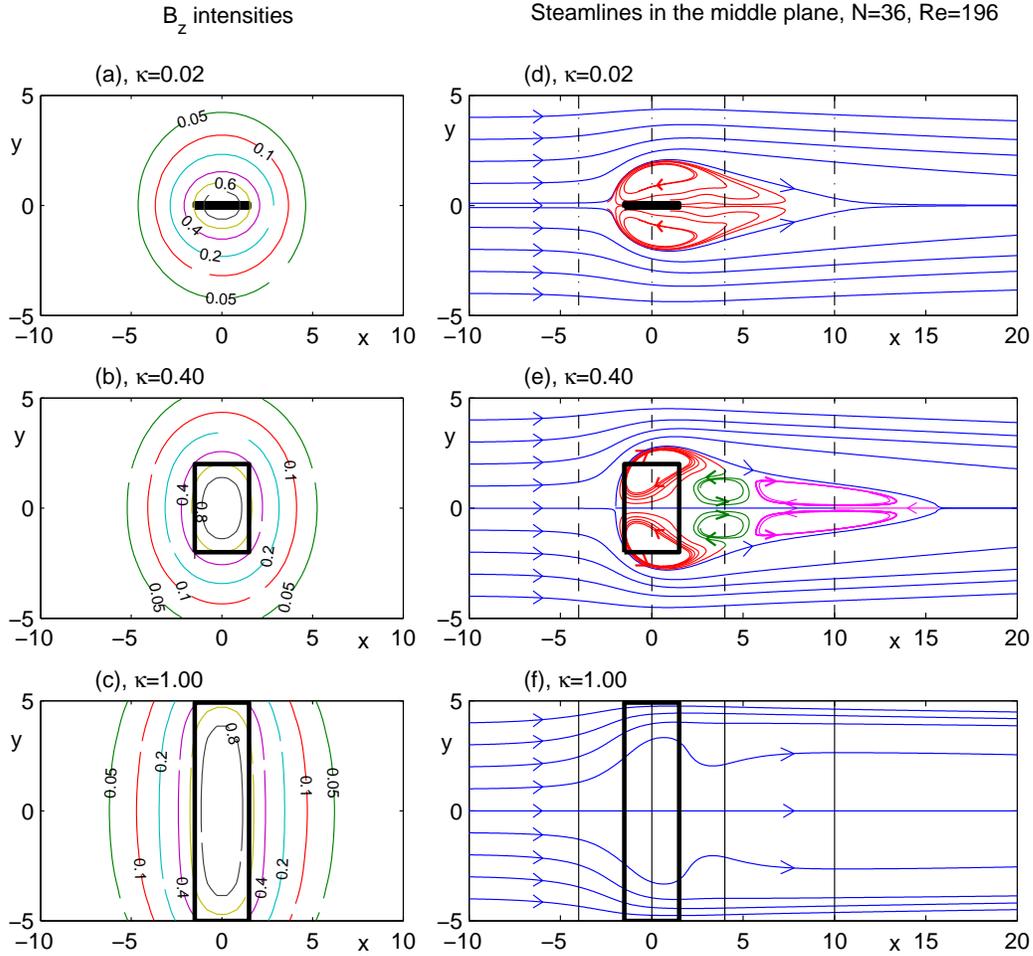}}
    \caption{\label{Fig:Streamlines}
    Contour lines of the transverse magnetic field component ($a$-$c$)
    and flow streamlines ($d-f$) in the middle plane ($z=0$) at
    $N=36$,$\;Re=196$ and $\kappa=0.02$ ($a,d$), 0.4($b,e$), and 1.0($c,f$). Magnetic pole
    is shown by bold lines. Dot-dashed($d$), dashed($e$) and solid($f$) vertical cuts
    denote the location of velocity profiles of Fig.~\ref{Fig:uvProfiles}.}
\end{figure}

In our opinion the most striking effect of spanwise heterogeneity
of the external magnetic field is shown in
Fig.~\ref{Fig:Streamlines}, where flow streamlines in the middle
plane, Fig.~\ref{Fig:Streamlines}($d-f$), are shown at the same
flow parameters, $N=36$ and $Re=196$. To get an impression about
the magnetic field configurations, the corresponding $B_z$ contour
lines are shown in Fig.~\ref{Fig:Streamlines}($a-c$). Depending on
the constrainment factor $\kappa$, one observes the following
stationary flow patterns: a vortex dipole for the magnetic blade
($\kappa=0.02$), Fig.~\ref{Fig:Streamlines}($a,d$); the stable
six-vortex ensemble for the middle magnet ($\kappa=0.40$),
Fig.~\ref{Fig:Streamlines}($b,e$); and no vortex motion for the
broad magnet ($\kappa=1.0$), Fig.~\ref{Fig:Streamlines}($c,f$).
The projection of the magnetic pole onto the middle plane is shown
by the bold solid line.

Let us qualitatively explain these flow patterns. The case of the
magnetic blade, Fig.~\ref{Fig:Streamlines}($a,d$), might be
roughly understood by considering the limiting case of a Lorentz
force which is pointwise in the spanwise direction. (By its
definition, ${\bf F_L = j \times B}$, the Lorentz force is a
volume force, however, if the distribution of heterogeneous
magnetic field $\bf B$ is very sharp in space, the Lorentz force
distribution is also very sharp.) As is known, see e.g.
\cite{Voropayev:Afanasyev:Book:1994}, such an instant retarding
force generates vorticity which results in two counter-rotating
vortices, a vortex dipole. Then, the induced vortices are advected
and diffused downwards from the source of the force.  In the case
of MHD flow, these two vortices are fixed by a sidelong gradient
of the magnetic field, so they stay on the place. We should note
that the similar recirculation has been obtained numerically
earlier by \cite{Gelfgat:Peterson:Sherbinin:1978}, and recently by
\cite{Cuevas:Smolentsev:Abdou:PRE:2006} for a creeping flow.

The case of the middle magnet, Fig.~\ref{Fig:Streamlines}($b,e$),
is explained briefly in the recent work of
\cite{Votyakov:PRL:2007} by means of a mutual interaction of
Lorentz and inertial forces. The first pair -- inner magnetic
vortices -- is an inheritor of the vortex dipole as in
Fig.~\ref{Fig:Streamlines}($d$); the third pair -- attached
vortices -- is of the same nature as a recirculation past a solid
body, Fig.~\ref{Fig:Introductory}($a$), while the intermediate
pair -- connecting vortices --  appear to make the coherent
rotation of the magnetic and attached vortices possible,
Fig.~\ref{Fig:Introductory}($b$), ~\ref{Fig:Streamlines}($e$). It
is clear why attached (hence, connecting) vortices are not induced
in the case of the magnetic blade: it has a well streamlined
shape, so there is no stagnancy region. This is in full analogy
with the flow around a solid body where an appearance of the
stagnancy region is strongly influenced by extent of streamlining.

Finally, the case of the broad magnet,
Fig.~\ref{Fig:Streamlines}($c,f$) shows no recirculation at given
parameters, because the flow pattern is influenced now mainly by
streamwise inward and outward gradients of the magnetic field. As
is known, see e.g. \cite{Shercliff:book:1962},
\cite{Kit:Peterson:Platnieks:1970}, \cite{Sterl:1990}, the
streamwise gradients and sidewalls of the duct are responsible all
together for a $M$-shaped profile of a streamwise velocity in the
spanwise direction. The $M$-shaped profile can also develop a
stagnant region in the middle of the duct at high interaction
parameter $N$, however, no recirculation has been discovered until
now. Moreover, as we shall prove in Section~\ref{sec:vorticity},
no recirculation is possible if an external magnetic field is
perfectly spanwise uniform. The present case of the broad magnet
shows a slight decline of $B_z$ moving towards sidewalls, see
Fig.~\ref{Fig:MF}($c$, solid line). In
Fig.~\ref{Fig:Streamlines}($c$), however, this decline is not
enough to develop magnetic vortices at the given interaction
parameter $N=36$. We will see later that the critical interaction
parameter $N_{c,m}$ which is needed to induce recirculation under
the broad ($\kappa=1.0$) magnetic gap is more than one hundred.

\subsubsection{Velocity profiles at different $\kappa$ and the same $N$ and $Re$. }

\begin{figure}
    \centerline{\includegraphics[width=13.5cm, angle=0,
    clip=yes]{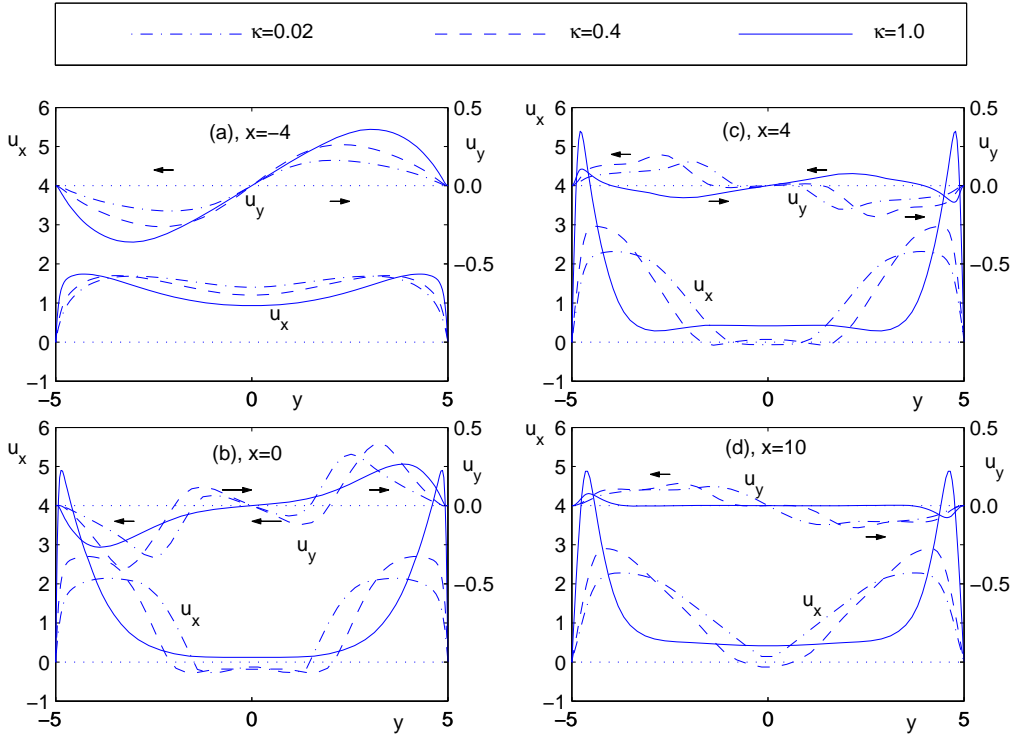}}
    \caption{\label{Fig:uvProfiles} Streamwise ($u_x$, three lower curves, left
    $y$-axis) and spanwise ($u_y$, three upper curves, right $y$-axis) velocities
    along spanwise cuts shown in Fig.~\ref{Fig:Streamlines}($d-f$). Middle plane, $z=0$,
    and $\;x=-4$($a$), $\;x=0$($b$), $\;x=4$($c$), $\;x=10$($d$).
    Constrainment factor $\kappa=0.02$ (dot-dashed), $\kappa=0.40$(dashed),
    and $\kappa=1.0$(solid lines), $N=36,\; Re=196$. Arrows show
    directions of spanwise redistribution of the flow.}
\end{figure}

Fig.~\ref{Fig:uvProfiles} shows quantitative data about  stream-
and spanwise velocity along different spanwise cuts. Different
line styles correspond  to the different $\kappa$'s. Arrows in
Fig.~\ref{Fig:uvProfiles} show directions of a spanwise
redistribution of the flow.

The largest braking Lorentz force, ${\bf F_L=j\times B}$, is
generated at the front of a magnetic obstacle where induced
electric currents ${\bf j}$ are maximum
% (see also Fig.~\ref{Fig:Middleplane}($d$) below)
and the magnetic field
${\bf B}$ is strong. This results in a deformation of the incoming
flow on the obstacle already at $x\!=\!-4$,
Fig.~\ref{Fig:uvProfiles}($a$). The deformation consists of
forming an inhomogeneous $M$-shaped profile of streamwise velocity
$u_x$ and an appearance of a spanwise flow, i.e. $u_y$, from the
center to sidewalls, as shown by arrows in
Fig.~\ref{Fig:uvProfiles}($a$).

Despite the order of magnitude difference in spanwise widths of
magnets for $\kappa=0.02$ (dot-dashed) and $\kappa=0.4$ (dashed),
their ranges of reverse velocity ($u_x\!\leq\!0$,
Fig.~\ref{Fig:uvProfiles}($b$)) along the $y$-axis differ only by
a factor one and half in the area of magnetic vortices. Positive
streamwise velocities of the vortices are smoothly transformed
into velocities of the external flow. Thus, for the wide range of
$\kappa$, the spanwise diameter of a single magnetic vortex
remains nearly constant provided the vortex is not influenced by a
sidewall. This is a manifestation of the fact that a decisive role
for the vortex is the width of the spanwise decline of a magnetic
field rather than the width of the magnet.

By forming magnetic vortices, the $M$-shaped profile of streamwise
velocity $u_x$ develops a negative value in the center, while the
spanwise velocity $u_y$ changes its sign twice. There is a
redistribution of the flow from a vortex inner side to the
centerline ($y=0$), and from an external side of the vortex to the
corresponding sidewall ($y=L_y$). On the contrary there are
distinct zones past the magnetic vortices where a  spanwise
redistribution is absent ($u_y\!=\!0$), see $|y|\!\leq\!0.8$ for
the magnetic blade ($\kappa=0.02$, dot-dashed) and
$|y|\!\leq\!1.5$ for the middle magnet (dashed),
Fig.~\ref{Fig:uvProfiles}($c$). Also, the zone of $u_y=0$ is
observed behind attached vortices, $|y|\!\leq\!1$,
Fig.~\ref{Fig:uvProfiles}($d$), $\kappa=0.4$.

Streamwise component of the velocity shows small changes with
rising $x$, \textit{cf.} $u_x$ in
Fig.~\ref{Fig:uvProfiles}($b-d$), except for a single peculiarity.
For the case of a broad magnet ($\kappa=1$, solid lines), one
observes that as $x$ increases, the maximum of $u_x$ is shifted
from sidewalls to the center due to diffusion of vorticity. In the
same time, the $u_y$ component changes its sign in the region near
sidewalls.

\subsection{Centerline profiles}\label{sec:centerlines}

In this Section, taking as an example the middle magnet,
$\kappa=0.4$, we show what forces are needed to induce vortices
inside and past a magnetic obstacle. To reach this goal we shall
analyze streamwise velocities along the centerline of the duct
($y=z=0$), i.e. centerline profiles.

\begin{figure}
    \centerline{\includegraphics[width=13.75cm, angle=0,
    clip=yes]{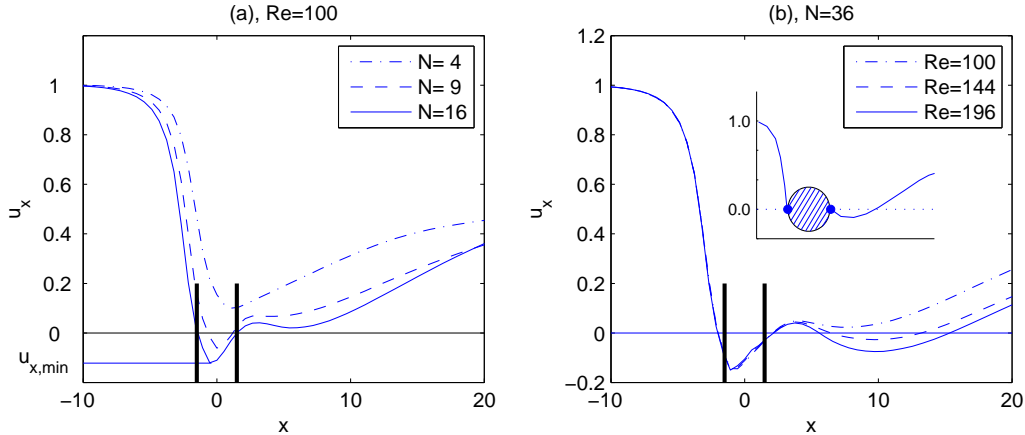}}
    \caption{\label{Fig:Centerlines} Streamwise velocity varied along
    the centerline $y=z=0$ for the middle magnet ($\kappa=0.4$):
    effect of $N$ at fixed $Re=100\;$ ($a$); and effect of $Re$ at fixed $N=36\;$
    ($b$). For ($a$): $N=4$(dot-dashed), 9(dashed), and 16(solid lines); for ($b$):
    $Re=100$(dot-dashed), 144(dashed), and 196(solid lines). Vertical bold lines show
    borders of the magnetic gap, $M_x=1.5$.
    Inset in ($b$) is a centerline velocity profile for a flow around a circular cylinder
    with attached vortices. The streamwise velocity is normalized to
    the centerline velocity of unretarding flow.}
\end{figure}

The main difference between a magnetic obstacle and a solid body
is the permeability of the obstacle depending on the retarding
Lorentz force, ${\bf F_L=j\times B}$. This braking force is the
largest in the center of the magnetic obstacle, where the magnetic
field ${\bf B}$ reaches the highest intensity. The characteristic
measure of the Lorentz force with respect to inertia is given by
the interaction parameter $N$: higher the $N$, the stronger is
${\bf F_L}$ and less penetrable is the space under the magnets.
Fig.~\ref{Fig:Centerlines}($a$) is to illustrate this behavior: a
dot-dashed line ($N=4$) shows that the streamwise velocity is
suppressed by approaching the magnet, however the braking force
for the given $N$ is not strong enough to reverse the flow. The
reversion happens at higher $N$, see dashed ($N=9$) and solid
($N=16$) lines being negative inside the obstacle. The lowest
velocity is marked in Fig.~\ref{Fig:Centerlines}($a$) by
$u_{\mbox{\scriptsize{x,min}}}$, it becomes zero at a critical
value $N_{c,m}$, and at $N\geq N_{c,m}$ one observes a
recirculation -- two inner magnetic vortices as shown in
Fig.~\ref{Fig:Streamlines}($d,e$). As the name points out, these
vortices belong completely to the MHD flow, and have nothing
common with a hydrodynamical flow around a cylinder. Rather these
vortices are similar to those appearing under the action of a
point braking force, see e.g. \cite{Afanasyev:2004}. The concrete
mechanism leading to the magnetic vortices is presented in Section
\ref{sec:vorticity}.

Now, we consider the behavior of centerline curves at fixed $N$
and varying $Re$,  Fig.~\ref{Fig:Centerlines}($b$) and apply the
analogy with ordinary hydrodynamics. Let us recall that the flow
around a solid cylinder shows a stagnant region with two attached
vortices when $Re$ is slightly higher than a critical value. The
typical centerline for this case is shown by inset of
Fig.~\ref{Fig:Centerlines}($b$). The same situation can happen for
the magnetic obstacle by increasing $Re$: the centerline curves
past the magnetic gap become negative again as shown in
Fig.~\ref{Fig:Centerlines}($b$). In fact, the attached vortices
are induced past the magnetic obstacle, see
Fig.~\ref{Fig:Streamlines}($e$, third pair of vortices) analogous
to those past a real solid body.

Note, that the first minimum in a centerline profile is almost
unperturbed when $Re$ increases at fixed $N$,
Fig.~\ref{Fig:Centerlines}($b$), and this is another strong
evidence that the vortices inside and past the magnetic obstacle
are of different physical origin. Since the interaction parameter
is given as $N=Ha^2/Re$, the magnetic vortices are enhanced by
decreasing the flow rate (i.e.~ $Re$). On the contrary, as follows
from Fig.~\ref{Fig:Centerlines}($b$), the attached vortices
manifest themselves by increasing the flow rate provided that the
intensity of the magnetic field (i.e.~$Ha$) as well as a spanwise
magnetic field gradient have already enforced  a reverse flow
inside the magnetic obstacle. In other words, there is a
qualitative distinction between magnetic and attached vortices:
the former arise when Re \textsl{decreases}, and the latter arise
when Re \textsl{increases} provided Ha is strong.

It is easy to see in Fig.~\ref{Fig:Streamlines}($e$) that magnetic
and attached vortices are co-rotating in the same direction
determined by the main flow movement. The only difference is the
driving torque: this is the Lorentz force for the magnetic
vortices, and the inertial force for the attached vortices.
Because the magnetohydrodynamic and attached vortices are
co-rotated, such a motion must be accompanied by a
counter-rotation which produces the intermediate pair of
connecting vortices, see Fig.~\ref{Fig:Streamlines}($e$, second
pair of vortices). The connecting vortices correspond to a local
maximum on centerline curves, Fig.~\ref{Fig:Centerlines}($b$)
behind the magnetic gap.

Thus, $N$ is responsible for the appearance of the magnetic
vortices, while $Re$ is responsible for the appearance of the
attached vortices. The connecting vortices are necessary for the
coherent rotation of the magnetic and attached vortices.

Three decades ago Gelfgat \textit{et~al.} have attempted to reveal
a kind of recirculation due to an external magnetic field by both
2D numerical simulation (\cite{Gelfgat:Peterson:Sherbinin:1978})
and physical experiments (\cite{Gelfgat:Olshanskii:1978}). They
saw a reverse flow numerically, and then designed a special
experiment which did not confirm the recirculation. As follows
from our results, the authors of the cited papers have not
realized that they observed and discussed different phenomena in
their numerical and experimental works. Their 2D numerical study
neglected an inertial term, what corresponds to a creeping flow,
and the observed reversion of the flow is just a sign of magnetic
vortices. Recently, similar 2D numerical work of
\cite{Cuevas:Smolentsev:Abdou:PRE:2006} showed the same effect in
a creeping flow without side walls. Thus, the numerical work of
Gelfgat precedes \cite{Cuevas:Smolentsev:Abdou:PRE:2006} by almost
thirty years.

The experimental report of \cite{Gelfgat:Olshanskii:1978} contains
a measured centerline profile (see Fig.7b in the cited work),
which falls behind the magnet, but does not approach negative
velocities. In the cited experiment, parameters are $N=7.5$ and
$Re=3.73\times 10^5$, so the authors noticed that the observed
falling velocity is due to inertial effects. Moreover, these
experiments were performed for the turbulent flow because $Re$ is
rather high. To obtain the desired recirculation, the authors
needed to just decrease further the inertial force by using a
lower flow rate. This would have resulted in a higher $N$ and, by
exceeding some threshold, made impossible the turbulent
pulsations. Then, the recirculation by the braking Lorentz force
might have been revealed with the appearance of magnetic and
attached vortices. Unfortunately, experiments with lower $Re$ have
not been performed, probably because the proper analysis of
centerline curves (see Fig.6) has been absent in those times.

\subsection{Existence regions of stationary flow patterns in parameter
space}\label{sec:diagram}

\begin{figure}
    \centerline{\includegraphics[width=13.75cm, angle=0, clip=no]{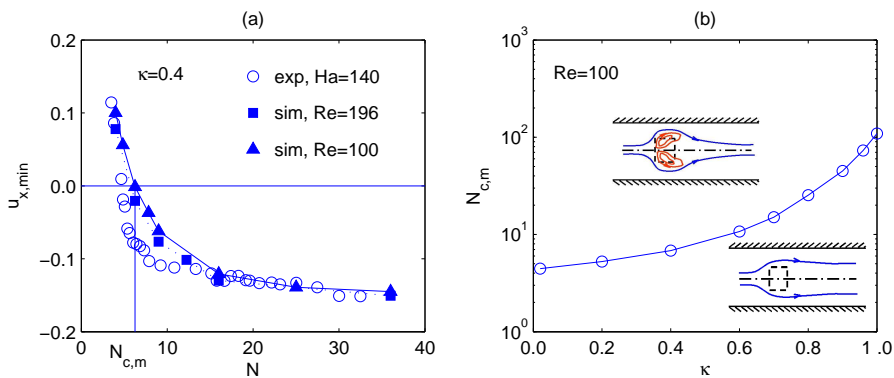}}
    \caption{\label{Fig:Diagram} Plot ($a$): minimum of streamwise velocity
    on the centerlines curves (see Fig.~\ref{Fig:Centerlines}$a$)
    depending on the interaction parameter $N=Ha^2/Re$ obtained
    in simulations (closed symbols) and in experiments (open circles)
    for $\kappa=0.4$ and $Ha=140$ (experiment), $Re=196$ (squares)
    and $Re=100$ (triangles). Plot ($b$): The critical interaction
    parameter $N_{c,m}$ depending on $\kappa$ for the appearance
    of magnetic vortices, 3d simulation, $Re=100$.}
\end{figure}

A sign of the recirculation enforced by the Lorentz force is the
negative value of $u_{\mbox{\scriptsize{x,min}}}$, see
Fig.~\ref{Fig:Centerlines}($a$). We have performed a series of 3D
simulations  for the different interaction parameters $N$ and
constrainment factors $\kappa$. An example of the series for the
middle magnet ($\kappa=0.4$) is shown in
Fig.\ref{Fig:Diagram}($a$). It is accompanied by experimental data
which were obtained by Oleg Andreev for the paper
\cite{Votyakov:PRL:2007} but not reported there due to lack of
space. Note, that the interaction parameter  $N=Ha^2/Re$ is
changed in experiments and numerics differently: experimentally by
varying the flow rate (i.e. $Re$) at fixed external magnets
(experimental $Ha=140$), while numerically  we kept $Re$ constant
and changed $Ha$. This is because of (i) experimental difficulties
to work with a low flow rate in the closed channel, and (ii)
numerical problems arising due to boundary layers at high Hartmann
numbers. Moreover, a transition to turbulent regime is another
factor that obscures stationary effects in numerics at high $Re$.

As one can see in Fig.~\ref{Fig:Diagram}($a$),
$u_{\mbox{\scriptsize{x,min}}}$ is positive at low $N$, and then
falls monotonically  to a constant level showing perfect agreement
between experiments and numerics for $N>15$. This excellent
agreement confirms that the flow under magnets depends solely on
$N$ and magnetic field configuration rather than $Re$ or inlet
profile, provided, the external magnetic field is strong enough to
suppress inward turbulent fluctuations. The latter condition was
satisfied in experiments by \cite{Andrejew:Kolesnikov:2006}. Also,
a similarity of the electric potential map at equal $N$ and
different $Re$ has been recently confirmed by 3D simulations and
comparison with experiments, see
\cite{Votyakov:Zienicke:FDMP:2006}. The slight difference shown in
Fig.~\ref{Fig:Diagram}($a$) at $N<15$ is explained by the fact
that the variation for $N$ has been defined differently in
experiments and numerics, as explained above.

Dependence of $u_{\mbox{\scriptsize{x,min}}}$ on $N$ becomes
negative at a critical value $N_{c,m}$. This value depends on the
constrainment factor $\kappa$. A series of 3D simulations has been
carried out for the range $0.02 \leq\kappa\leq 1$ in the vicinity
of $u_{\mbox{\scriptsize{x,min}}}(N) \approx 0$ and the results
are given in Fig.~\ref{Fig:Diagram}($b$) which shows the
separation line between regions for stable flow without and with
magnetic vortices\footnote{It is tempting to call
Fig.~\ref{Fig:Diagram}($b$) a stability diagram. We renounce from
doing this, because --- strictly speaking --- it is not proven
that the change of topology of the flow patterns is caused by a
change of stability. We find this assumption highly probable, but
we did not compute eigenvalues of the Jacobi matrices for the
stable flow patterns to check whether the transition is a
bifurcation or not (see also our remark at the end of the
introduction). We continue to call $N_{c,m}$ the critical
interaction parameter. This has to be understood not in the sense
of a change of stability, but instead in the sense of separating
stable solutions with a different topology.}. During this series
of 3D simulations we did never find a hint that both flow patterns
--- the one with and the one without magnetic vortices --- coexist
for the same pair of parameters, i.e. both solutions are stable,
but have different basins of attraction. However, we have checked
different initial conditions only for some parameter combinations
and did not carry out a systematic search with different initial
conditions. Therefore, we can not conclude that the case of
coexistence of two solutions is not possible.

From the separation line of Fig.~\ref{Fig:Diagram}($b$), as we
have it numerically determined, the following trends are visible:
lower the $\kappa$, smaller the influence of sidewalls; the case
of $\kappa\rightarrow 0$ corresponds to a free flow. Larger the
$\kappa$, more uniform is the braking Lorentz force in the
spanwise direction. Therefore, in order to induce inner vortices
at larger $\kappa$ it is necessary to apply  a larger critical
interaction parameter $N_{c,m}$. For middle magnets, $\kappa\!
\leq \! 0.5$, the critical value of $N_{c,m}$ is of the same order
of magnitude, $N_{c,m}\!\approx\! 6$, and for broad magnets,
$\kappa \geq 0.8$, it increases up to $N_{c,m}\!=\!109$ at
$\kappa\!=\!1$. The latter case has been earlier discussed  in
Fig.~\ref{Fig:Streamlines}($c,f$) for $N\!=\!36$ as an example of
a vortex-free flow pattern. We prove in next Section that any
recirculation is impossible if the external magnetic field is
perfectly spanwise uniform.

\subsection{The mechanism to induce recirculation.
    Vorticity, electric field, drop of pressure.}\label{sec:vorticity}

One has to analyze the electric field inside the magnetic obstacle
for different $\kappa$ in order to understand why the appearance
of vortex motion is strongly dependent on the spanwise variation
of the external magnetic field.

Let us recall that the electric potential, $\phi$, is distributed
according to the Poisson equation, $\Delta \phi=\nabla\cdot({\bf
u\times B}) ={\bf B \cdot \textbf{\textit{w}}}$, where
${\textbf{\textit{w}}}=\nabla\times{\bf u}$ is the vorticity. Note
that the projection of ${\textbf{\textit{w}}}$ on the externally
fixed vector ${\bf B}$ plays a role of the induced electric charge
density which creates the electrostatic field ${\bf
E}=-\nabla\phi$. It is possible to derive the direction of the
field ${\bf E}$ from the knowledge of positive and negative
extrema in the ${\bf B \cdot \textbf{\textit{w}}}$ distribution,
by roughly assuming that these extrema can be approximated as
point charges. As shown in electrostatics, the maximum (minimum)
of rhs in Poisson equation, i.e. $B_z\omega_z$, creates the
minimum (maximum) in $\phi$ distribution.

\begin{figure}
    \centerline{\includegraphics[width=13.5cm, angle=0,
    clip=yes]{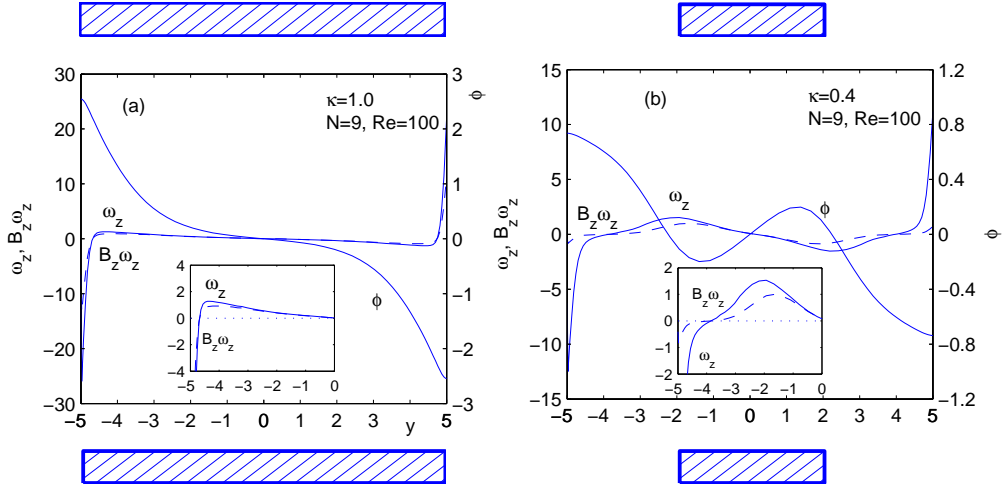}}
    \caption{ \label{Fig:VorticityElField}
    Spanwise cuts, $x\!=\!z\!=\!0$, of vorticity, $\omega_z$ (left axis, solid),
    magnetic field multiplied by vorticity, $B_z\omega_z$ (left axis, dashed), and
    electric potential,$\phi$ (right axis), for $\kappa=1$(plot $a$) and
    $0.4$(plot $b$); $N=9$, $Re=100$.
%    At sidewalls: $\omega_z(y\!=\!\pm
%    L_y)\!\approx\!\pm\!25$($a$), $\pm\!85$ ($b$), and $B_z\omega_z(y=\pm L_y)\approx
%    \pm 1.75$($a$), $\pm 50$($b$).
    Filled rectangles at top and bottom show schematically magnets.}
\end{figure}

For definiteness we consider the central spanwise cut,
$x\!=\!z\!=\!0$, where $B_x=B_y=0,\; B_z\geq 0$, and ${\bf B \cdot
\textbf{\textit{w}}}\!=\!B_z\omega_z\!\approx\! -B_z\partial
u_x/\partial y$. This cut goes through the center of the magnetic
gap and most expressively develops peculiarities typical of any
spanwise cut. Vorticity $\omega_z$, the product of $B_z \omega_z$,
and resulting electric potential $\phi$ along this central
spanwise cut are shown in Fig.~\ref{Fig:VorticityElField} for the
broad ($a$) and middle ($b$) magnet. The magnets are depicted on
the top and bottom by filled rectangles.

The behavior of vorticity, $\omega_z\!\approx\! -\partial
u_x/\partial y$, can be easily understood from the spanwise
deformation of the streamwise velocity, i.e. $M$-shaped velocity
profile, Fig.~\ref{Fig:uvProfiles}($c$). Such a profile is
characterized by two jets sidelong streamlining the magnetic
obstacle, and each jet has the internal and external slope, hence,
the vorticity alternates its sign by passing the velocity maximum
of the jet. The steepness of the external side of the jet is
significantly higher than that of the internal side, especially
for the larger $\kappa$ because the external slope is adjoined
with no-slip sidewalls. Fig.~\ref{Fig:VorticityElField} shows that
the sidewalls vorticity, i.e. the external slope vorticity, is
larger by an order of magnitude than the vorticity adjoining the
center, i.e. the internal slope vorticity. To stress this fact,
Fig.~\ref{Fig:VorticityElField} shows both global and zoomed
(insets) views for $\omega_z$ and $B_z\omega_z$.

When the spanwise variation of the magnetic field $B_z(y)$ is
weak, e.g. for the broad magnet ($\kappa=1$) shown in
Fig.~\ref{Fig:VorticityElField}($a$), the product of $B_z\omega_z$
reflects completely the quantitative difference in $\omega_z$
taken on the internal and external jet sides. Because the external
$\omega_z$ is much higher than the internal $\omega_z$, the
distribution of $\phi$ is determined mainly by the vorticity
generated on sidewalls rather than by that produced inside the
obstacle. Therefore, the negative (positive) external $\omega_z$
induces strong positive (negative) potential on the sidewalls, and
 $\phi(y)$ drops monotonically along the $y$-axis,
i.e. the spanwise electrostatic field $E_y =-\partial\phi/\partial
y$ is always positive and does not change its sign between
sidewalls.

If the magnetic field is perfectly uniform along the $y$-axis, the
electric potential distribution is completely governed by the
sidewalls vorticity, and the alternating vorticity on the internal
slopes of the $M$-shaped profile does not contribute to $\phi(y)$
even at large interaction parameter $N$. As a result, the
electrostatic field ${\bf E} \approx (0,E_y,0)$ in the region of
magnetic obstacle is always directed opposite to electromotive
force, ${\bf u\times B} \approx (0,-u_xB_z,0)$. By braking the
flow, streamwise velocity $u_x$ and spanwise field $E_y$
accordingly approach zero by keeping their signs. A stagnant
region can develop where streamwise velocity is small but still
positive because the electric field is always in the same
direction. Therefore, recirculation is impossible.

Let us consider the case when the magnetic field intensity,
$B_z(y)$, drops from the center to sidewalls. Here, the high
vorticity generated by sidewalls in the product of $B_z\omega_z$
is suppressed by a low intensity of the magnetic field. This is
shown by the middle magnet ($\kappa=0.4$),
Fig.~\ref{Fig:VorticityElField}($b$)),  the corresponding $B_z(y)$
is presented by dashed lines in Fig.~\ref{Fig:MF}($c$). The
product of $B_z\omega_z$ resembles the  $\omega_z(y)$ behavior
inside the magnetic obstacle, since there $B_z(y) \approx
\mbox{const}$, i.e. the internal vorticity is well presented in
$B_z\omega_z$. Outside the obstacle, $B_z(y)$ rapidly decreases,
so the external vorticity is remarkably weakened: as follows from
Fig.~\ref{Fig:VorticityElField}($b$), the dashed $B_z\omega_z$
curve diverges from the solid $\omega_z$ curve. The largest
difference between $\omega_z$ and $B_z\omega_z$ are present on
sidewalls where the external vorticity is very high due to no slip
boundary conditions, while $B_z\omega_z$ at $|y|=L_y$ is of the
same magnitude because its alternating extrema at $|y|\approx M_y$
correspond with borders of the magnet. Therefore, the dependence
of $\phi(y)$ is allowed to be influenced by the internal slopes of
the $M$-shaped profile, which produces extrema on $\phi(y)$ at
$|y|\approx M_y$ alternating with the sign of $\phi$ on sidewalls.
These internal extrema on $\phi(y)$ reverse the spanwise electric
field i.e. $E_y =-\partial\phi/\partial y$ becomes negative inside
a finite magnetic obstacle.

The reverse spanwise electrical field appearing for spanwise
decaying magnetic field is a necessary but not sufficient
condition for the appearance of reverse flow. It is a necessary
condition, because otherwise no Lorentz force pointing in negative
$x$-direction would appear. This becomes clear looking at
equations (\ref{eq:NSE:momentum}) and (\ref{eq:NSE:Ohm}) for
Lorentz force and current density: ${\bf j}\times{\bf B}=-\nabla
\phi \times {\bf B} -B^2{\bf u}+({\bf u}\cdot{\bf B}){\bf B}$.
Considering again the central spanwise cut, $x=z=0$, where ${\bf
B}=(0,0,B_z)$, ${\bf u}\approx (u_x,0,0)$ and ${\bf j}\approx
(0,j_y,0)$, one can write the $x$-component of the Lorentz force
as follows: $({\bf j}\times{\bf B})_x \approx
j_yB_z=E_yB_z-B_z^2u_x$. The third term has vanished, the second
term corresponds to a frictional force proportional to the actual
velocity. The only term, which can provide a driving force for the
reverse flow is the first term, and this only then when $E_y$ is
negative, i.e. when a reverse spanwise electrical field exists.
Only the existence of a reverse spanwise electrical field is not
sufficient to drive reverse flow, because it has to reach a
minimal strength. This becomes clear from the $y$-component of
Ohm's law for the electrical current density at the central
spanwise cut: $j_y=E_y-u_xB_z$. To see this, one needs an
additional information concerning the global behavior of the
electrical current in our system. The electrical current is
organized in two horizontal loops, with a strong negative spanwise
current (i.e. $j_y<0$) directly under the magnet, which divides at
the sidewalls into one loop before the magnet and the other behind
the magnet (see figures 16, 17 and 18 of our former paper
\cite{Votyakov:Zienicke:FDMP:2006}). The sign of $j_y$ does not
change, when the interaction parameter is increased. When $E_y$
just changes sign at a certain value of $N$ at the start of
electric field reversal, $-u_xB_z$ has to be negative. This is
only possible for (still) positive $u_x$, and consequently $N$ has
to be further increased until the reverse spanwise electrical
field becomes so strong that it equals the value of $j_y$. This
characterizes, in fact, the critical $N_{c,m}$, for which the
velocity $u_x$ is exactly zero. Let us denote this value of the
reverse spanwise electrical field by $E_{y,c}$. A further increase
of the interaction parameter makes $E_y$ stronger negative than
$j_y$ with the consequence that $u_x$ has to be smaller than zero,
which is equivalent to the existence of reverse flow and
recirculation. Thus, summing all up, the role of  spanwise
magnetic field gradient is to suppress the external vorticity and
promote the internal vorticity in the product of ${\bf B
\cdot\omega}$. When by a further increase of the interaction
parameter the internal spanwise electrostatic field is reversed
stronger than the critical $E_{y,c}$, the recirculation appears.

\begin{figure}
    \centerline{\includegraphics[width=13.75cm, angle=0,
    clip=yes]{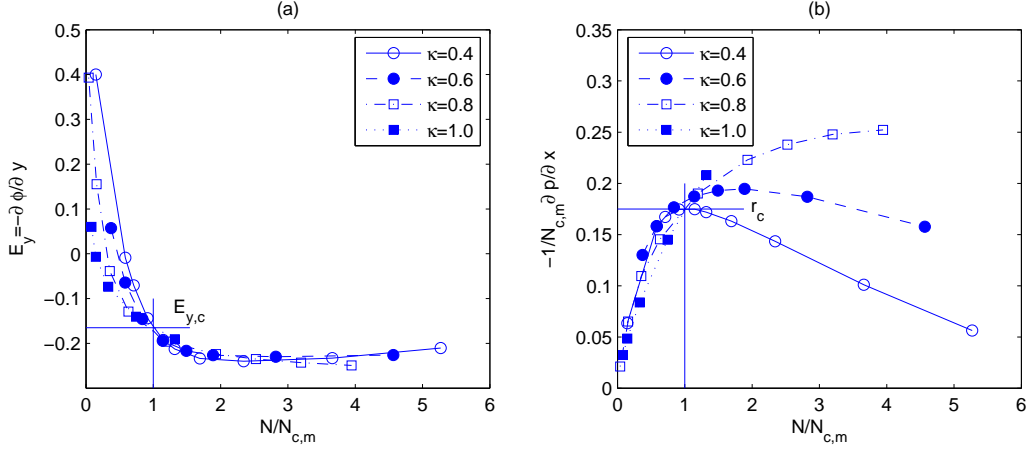}}
    \caption{\label{Fig:ElfieldDragforce} For various magnetic
    field configurations, $\kappa=0.4$ (open circles), 0.6
    (filled circles), 0.8 (open squares), 1.0 (filled squares),
    the dependencies on  $N/N_{c,m}$  in the center of magnetic obstacle,
    $x\!=\!y\!=\!z\!=\!0$ of the spanwise electrostatic field, $E_y\!=\!-\partial\phi/\partial y$,
    (plot $a$), and the normalized resistance to the flow, $r=(-\partial p/\partial
    x)/N_{c,m}$, (plot $b$).}
\end{figure}

Fig.~\ref{Fig:ElfieldDragforce}($a$) shows $E_y
=-\partial\phi/\partial y$ for the different magnets taken in the
center of the  magnetic gap, $x\!=\!y\!=\!z\!=\!0$, as a function
of the interaction parameter $N$ normalized with the critical
value $N_{c,m}$. Letting the appearance of recirculation be the
referred flow pattern, the ratio $N/N_{c,m}$ characterizes the
power of recirculation for various $\kappa$. One can see that the
curves are close to each other when $N/N_{c,m} \gtrsim 1$.
Moreover, by taking $N/N_{c,m}=1$ one finds that the critical
$E_{y,c}\approx -0.17$  is independent of $\kappa$. That is, the
critical magnitude of the spanwise electrostatic field, $E_{y,c}$
is a universal parameter which is the same for different magnetic
field configurations. This fact is clear because $B_z=1$ and
$j_y=E_{y,c}$ when the recirculation starts, i.e.
$u_x\!=\!u_y\!=\!0$. So, $E_{y,c}$ is indeed the critical braking
Lorentz force because $F_{L,x}=j_yB_z=E_{y,c}$.

%If one considers the $x$-component of Navier-Stokes momentum
%equation~(\ref{eq:NSE:momentum}) in the place $(x,0,0)$ of the
%originating recirculation, i.e. \mbox{ and } B_z=1$, inertial and
%viscous forces are neglected, one obtains  $\partial p/\partial x
%+ N_{c,m}E_{y,c}=0$.

Let $r\!=\!-(\partial p/\partial x)/N_{c,m}$ at
$x\!=\!y\!=\!z\!=\!0$ be the resistance to the flow inside the
magnetic obstacle. At the beginning of recirculation the
Navier-Stokes equation is simplified up to $\partial p/\partial x
+ N_{c,m}E_{y,c}=0$, so $r_c=-E_{y,c}$. The behavior of $r$ as a
function of $N/N_{c,m}$ for various $\kappa$ is shown in
Fig~\ref{Fig:ElfieldDragforce}($b$). One can see that the change
of the flow regime is accompanied by the change of the slope  in
the resistance of $r(N/N_{c,m})$. For those $\kappa$ where the
sidewall influence is insignificant, the appearance of
recirculation at $N/N_{c,m}\gtrsim 1$ results in a drop of the
resistance despite the fact that $N$ increases.

The second effect of a drop in the resistance due to magnetic
recirculation has the same explanation as the drag crisis well
known in  ordinary hydrodynamics. It appears in the wake of a
circular cylinder when the boundary layer on the cylinder surface
undergoes a transition from the laminar to turbulent mode. It
causes a substantial reduction in the drag force analogous to that
in the wake of the magnetic obstacle.

\subsection{Similarity of MHD flows inside the magnetic obstacle}

As follows from Eq.(\ref{eq:NSE:momentum}) the viscous force
$\Delta {\bf u}$ is scaled by Reynolds number $Re$, therefore at
high $Re$ and far from the walls it plays a minor role. As a
result the flow must be governed by the interaction parameter $N$
defining  the ratio between magnetic and inertial forces.  This
peculiarity was marked already in
\cite{Votyakov:Zienicke:FDMP:2006} by comparing experimental and
numerical electric potential distributions found at different $Re$
and similar $N$.

\begin{figure}
    \centerline{\includegraphics[width=13.5cm, angle=0,
    clip=yes]{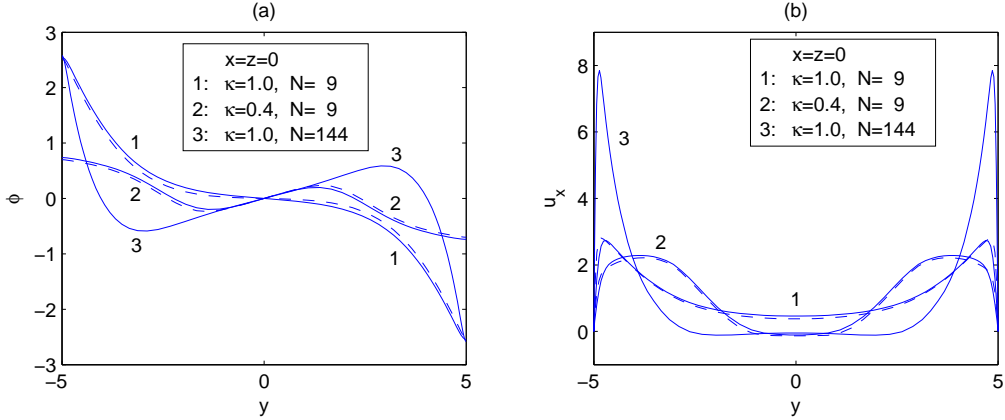}}
    \caption{ \label{Fig:Similarity}
    Spanwise dependence ($x\!=\!z\!=\! 0$) of electric
    potential (plot $a$) and streamwise
    velocity (plot $b$) for the broad
    ($\kappa=1.0,\; N_{c_m}=109$, solid and dashed curves $1,3$)
    and middle
    ($\kappa=0.4,\; N_{c_m}=6.8$, solid and dashed curves $2$) magnetic obstacle.
    $N=9$ (curves $1,2$) and 144 (curves $3$), $Re=100$ (solid) and 400 (dashed lines).
    Notice that for the curves $2$ and $3$ the constrainment factors $\kappa$ are different,
    while the relations $N/N_{c,m}\approx 1.32$  are equal.}
\end{figure}

In the present paper, Fig.~\ref{Fig:Similarity} demonstrates  the
same behavior for the broad (curves $1$ and $3$) and middle
(curves $2$) magnet. Solid and dashed curves $1,2$ have the same
$N$ but $Re=100$ for solid and $Re=400$ for dashed lines.

It is expected after discussing Fig.~\ref{Fig:ElfieldDragforce}
 that the results given in Fig.~\ref{Fig:Similarity}
are similar even  for different constrainment factors $\kappa$ and
interaction parameters $N$ provided the corresponding ratios
$N/N_{c,m}$ are equal. Curve 2 for the middle magnet is plotted
for $N=9$, i.e. $N/N_{c,m}(\kappa=0.4)=9/6.8=1.32$ and curve $3$
for the broad magnet -- for $N=144$, i.e.
$N/N_{c,m}(\kappa=1.0)=144/109=1.32$. One can see that in the
central part $|y| \lesssim 2$ being approximately the spanwise
dimension of the less magnet ($M_y=0.4\times 5=2$) both electric
potential ($a$) and streamwise velocity ($b$) are also close to
each other,  \textsl{cf.} curves $2$ and $3$.

%
%\subsection{Pressure, vorticity, electric potential and electric currents of the middle plane}
%
%\begin{figure}
%    \centerline{\includegraphics[width=13.5cm, angle=0,
%    clip=yes]{Middleplane.eps}}
%    \caption{ \label{Fig:Middleplane}
%    Pressure ($a$), $z$-component of vorticity ($b$),
%    electric potential ($c$) and electric currents ($d$)
%    of the middle plane, $z=0$, $\kappa=0.4$, $N=36$, $Re=196$.
%    Bold rectangle shows the borders
%    of magnetic poles. For ($b$) and ($c$) negative contour levels are given  by
%    dashed lines.}
%\end{figure}
%
%\karaul{Yuri, please write more here Fig.~\ref{Fig:Middleplane},
%0.5-1.0 page  .} \vspace{15cm}
%

\subsection{3D peculiarities of the flow} \label{sec:3dresults}

\begin{figure}
    \centerline{\includegraphics[width=13.5cm, angle=0,
    clip=yes]{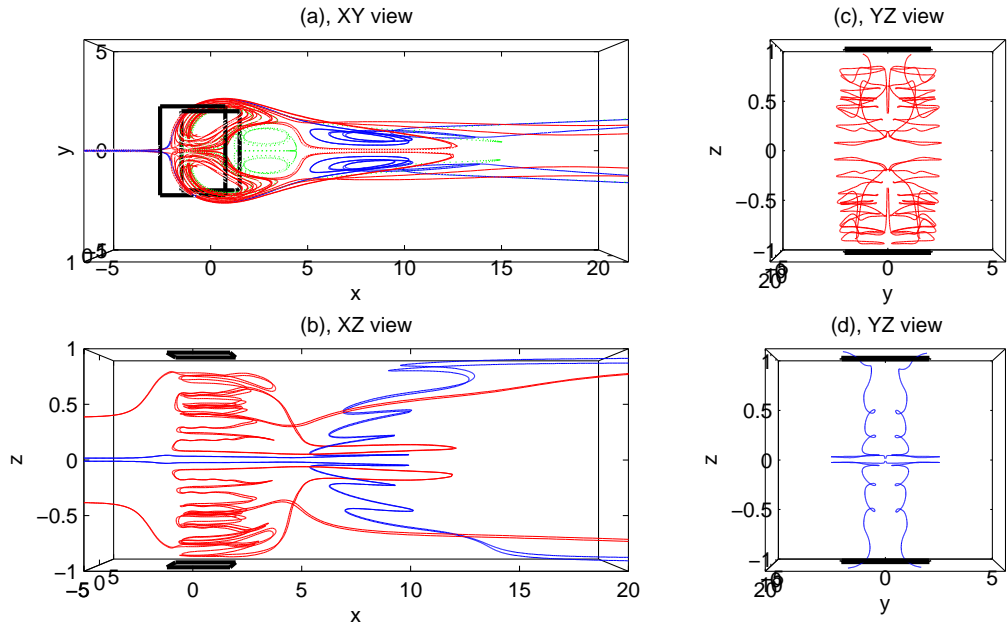}}
    \caption{ \label{Fig:Stream3D}
    Perspective projection of 3D vortex disposition  for middle magnet and $N\!=\!16$, $Re\!=\!196$:
    $XY$ view, magnetic, connecting and attached vortices ($a$),
    $XZ$ view, magnetic and attached vortices ($b$),
    $YZ$ view, magnetic vortices ($c$),
    $YZ$ view, attached vortices ($d$).}
\end{figure}

Fig.~\ref{Fig:Stream3D} shows different 3D space perspective views
of the six-fold vortex pattern for the middle magnet
($\kappa=0.4$), $N=16$, and $Re=196$. For easy visualization, plot
$a$ shows the $XY$ projection of all vortices, i.e. magnetic,
connecting, and attached, while other plots present only part of
the recirculation. Plot $b$ shows the $XZ$ projection for magnetic
and attached vortices, plots ($c-d$) are both $YZ$ projections of
the magnetic, and attached vortices, correspondingly.

The following 3D features are observed in Fig.~\ref{Fig:Stream3D}:
(i) a helical motion inside every vortex; (ii) the axes of
rotation of the magnetic vortices are parallel to the lines of
external magnetic field;  (iii) the two magnetic vortices are
closely adjoined to each other and taken together they form a
barrel extending in its central part and located mainly between
magnetic poles; (iv) two helices of attached vortices are not
adjoined and are arched along $x$-direction.

\begin{figure}
    \centerline{\includegraphics[width=13.5cm, angle=0,
    clip=yes]{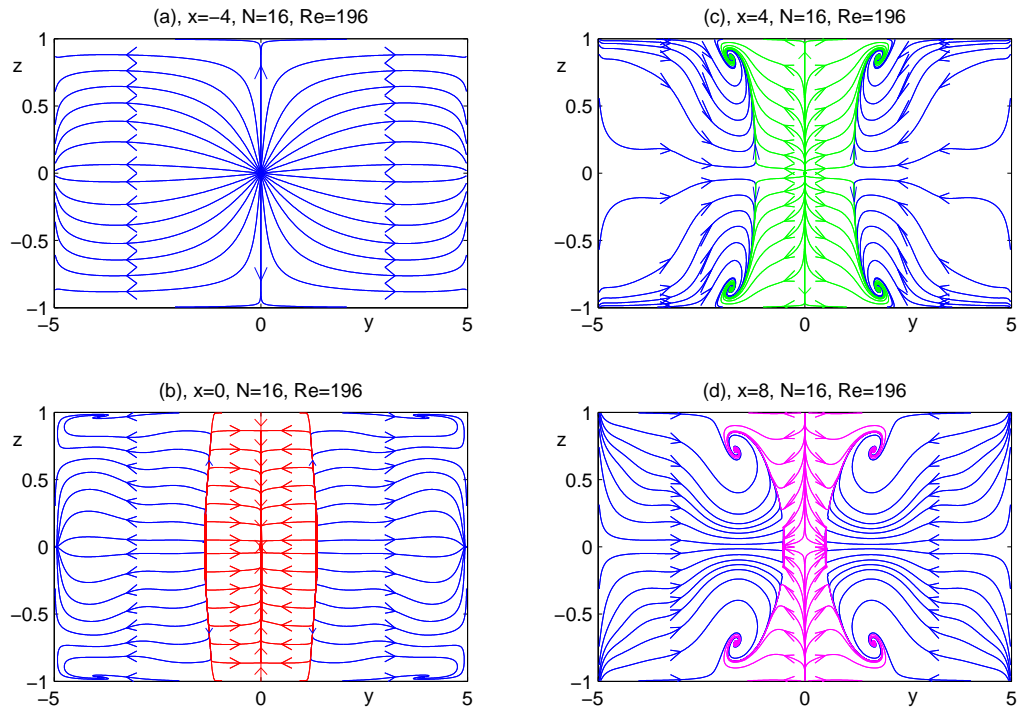}}
    \caption{ \label{Fig:StreamYZ} Flow streamlines constructed from $u_y$,
    $u_z$ velocity components in vertical slices $x=-4(a)$,
    0($b$), 4($c$), and 8($d$);  $N\!=\!16$, $Re\!=\!196$.}
\end{figure}

Additionally, Fig.~\ref{Fig:StreamYZ} present flow streamlines
constructed from $u_y$ and $u_z$ components of velocity field for
few characteristic vertical slices: at front of the magnet
($x\!=\!-4$, plot $a$), and the vertical cuts of magnetic
($x\!=\!0$, plot $b$), connecting ($x\!=\!4$, plot $c$) and
attached ($x\!=\!8$, plot $d$) vortices. These ($u_y$, $u_z$)
streamlines are not real moving lines of fluid particles in the
vertical slices, because the particles have also the $u_x$
velocity component. Instead, these streamlines are to show
ascending and descending paths of fluid particles in vertical
slices as projection.

At $x=-4$, Fig.~\ref{Fig:StreamYZ}$a$,  one observes the braking
effect of the Lorentz force making the flow streamlines diverge
from the central point $y=z=0$, where the greatest change of
$\partial u_x/\partial x=-(\partial u_y/\partial y+\partial
u_z/\partial z)$ occurs. Because the $u_x$ component is not
involved in the plot, this effect looks as the source of the flow.
Also there is a motion in $z$ direction caused by a tendency to
form Hartmann layers, see \cite{Votyakov:Zienicke:FDMP:2006}.

At $x=0$, Fig.~\ref{Fig:StreamYZ}$b$, one firstly notices the
abrupt change of the motion in $y$ direction at $|y|\approx 2$.
This helps us to see the vertical borders of the magnetic vortices
where the $u_y$ reverse its sign. Moreover, a barrel shape of the
magnetic vortex dipole can be clearly seen. Directions of arrows
on the flow streamline demonstrate a drift of the flow from
top/bottom walls towards the center inside the magnetic vortices,
and a drift in the opposite direction outside. Such a drift is a
reflection of helical motion which can be seen in perspective in
Fig.~\ref{Fig:Stream3D}.

Similar to Fig.~\ref{Fig:StreamYZ}$b$, sharp borders between the
main flow and connecting ($x=4$, Fig.~\ref{Fig:StreamYZ}$c$) and
attached ($x=8$, Fig.~\ref{Fig:StreamYZ}$d$) vortices are
observed. One also observes  a remarkable vertical drift to and
from the top and bottom walls depending on the direction of
horizontal rotation. The vertical drift at $x=4$ and $x=8$ is more
pronounced than that of at $x=0$ because the magnetic field has a
much larger intensity in the case of $x=0$. The new phenomena
compared to Fig.~\ref{Fig:StreamYZ}$b$ are swirls in the corners
of the duct shown in plots ($c,d$). Similar swirls can also appear
when a MHD flow has no recirculation. They arise due to the
inertial force and destruction of Hartmann layers resulting from a
decline of the magnetic field, see
\cite{Votyakov:Zienicke:FDMP:2006}.

The vertical drifts shown in Fig.~\ref{Fig:StreamYZ}$b-d$ are
caused by rotation of the fluid and can be understood as a
manifestation of the hydrodynamic Ekman pumping effect, see e.g.
\cite{Davidson:book:2001}. There are six rotating columns in the
MHD flow studied here, and each one has its own primary horizontal
motion and secondary vertical drift. Taken all together it gives
rise to the helical motion clearly observed  in
Fig.~\ref{Fig:Stream3D}.

The whole 3D space trajectory for the infinitesimally small volume
of the fluid -- a fluid particle -- can now be  described in the
following way. Far upstream from the magnetic system, the particle
moves straight under the pressure gradient. Approaching the region
of influence of the magnetic obstacle, the particle turns aside
towards the closest corner due to the action of the braking
Lorentz force and reaches a boundary layer. If the particle is not
captured by recirculation, it then passes the region of the
magnetic obstacle in the bulk of two jets sidealong streamlined
the obstacle. If the particle is captured by recirculation, it
first goes down (up) from the top (bottom) towards the middle
plane in the helix of the magnetic vortex. In the middle plane,
this trajectory is taken by the helix of the closest connecting
vortex and passes helically towards the top (bottom) wall to come
again to the boundary layer and dissipate the kinetic energy.
Then, the particle can be caught by the helix of the attached
vortex and be drifted slowly to the middle plane where it finally
becomes free to go downstream from the magnet.

Recirculation under magnetic poles brings new details into top and
bottom boundary layers perpendicular to the magnetic field. These
layers are Hartmann layers in the case of a constant magnetic
field, Ekman layers in the case of an axisymmetric rotating flow
bounded by a fixed horizontal plate, and Ekman-Hartmann layers
when the axisymetric rotating flow is subject to a constant
vertical magnetic field, \cite{Acheson:Hide:1973},
\cite{Desjardins:Dormy:Grenier:1999}\footnote{We thank the Referee
who brought these works to our attention.}. We checked velocities
profiles $u_x(z)$ in the region of magnetic vortices and found no
satisfactory agreement with the current theory for Ekman-Hartman
rotating flows. The reasons  for this disagreement are probably
that the recirculation induced by a heterogenous external magnetic
field shows details which are not compatible with the present
analytic theory: (i) the rotating flow is not axisymmetric, (ii)
neither magnetic field nor angular rotation are constant, (iii)
this is a system of six rotating flows. The quantitative analysis
of these layers requires further detailed investigation. At the
moment we can conclude only that these layers are important to
stabilize recirculation even at high Reynolds numbers as shown in
next Section.

\subsection{Recirculation in a 3d flow versus a vortex
generation in a 2D flow} \label{sec:3dversus2d}

This paper is devoted to a stationary 3D MHD flow. Recently,
\cite{Cuevas:Smolentsev:Abdou:2006} have reported a vortex
generation in a 2D flow induced by a magnetic obstacle. In this
Section, we discuss how their 2D results are related to our 3D
results.

In our opinion, it is an open question on how to build a 2D model
at high $Re$. Two-dimensionality assumes that the flow rate is
kept constant in the plane under consideration. This assumption is
certainly wrong in the case of a local magnetic field, where
Hartmann layers are formed (destroyed) under inward (outward)
magnetic field gradient, i.e. streamwise velocity profile in the
transverse direction is becoming more (less) flat, therefore the
fluid must go out of  (go into) the plane, see discussion about
Fig.8,9 given by \cite{Votyakov:Zienicke:FDMP:2006}.

On one hand, the friction imposed by a no-slip wall stabilizes a
flow because it provides a sink for kinetic energy. For instance,
in ordinary hydrodynamics there are numerous experiments which
illustrate that the confinement of the endplates increases the
stability of the wake, see e.g. \cite{Shair:Grove:etal:1963},
\cite{Nishioka:Sato:1974}, \cite{Gerich:Eckelmann:1982},
\cite{Lee:Budwig:1991}. These examples include the delay of the
critical Reynolds number for vortex shedding and the extension of
the Reynolds number range for a 2D laminar shedding.

On the other hand, the case of a MHD flow under a local magnetic
field always requires to take into account top and bottom
endplates because these plates carry magnetic poles. In reality,
one can increase the distance between poles to have more
two-dimensionality in the middle plane, but this automatically
decreases the degree of space heterogeneity of the external
magnetic field. So, it becomes an issue whether it is practically
possible to design a strongly heterogenous magnetic field having a
large distance between magnetic poles.

There is a method to take no-slip top/bottom plates into
consideration by means  of the Hartmann friction term, see e.g.
\cite{Lavrentiev:Molokov:Magnetohydrodynamics:1990}. This
assumption has been used also by
\cite{Cuevas:Smolentsev:Abdou:2006}. However, this averaged
approach is well validated only in the case of a small flow rate,
i.e.  low $Re$, even when the magnetic field is not strongly
varying. It follows from the fact that past the magnetic obstacle
the Hartmann friction takes a form of the Hele-Shaw friction based
on the assumption that velocity is parabolic along the $z$-axis.
This is well validated only for viscous flows where the vorticity
advection does not play a significant role, \cite{Riegels:1938}.
The formal vertical Hele-Shaw friction term inserted into the 2D
Navier-Stokes equations does not describe quantitatively the
behavior of the system with high $Re$. We believe that this
friction has no significant influence when the advection is
strong.

\begin{figure}
    \centerline{\includegraphics[width=13.5cm, angle=0,
    clip=yes]{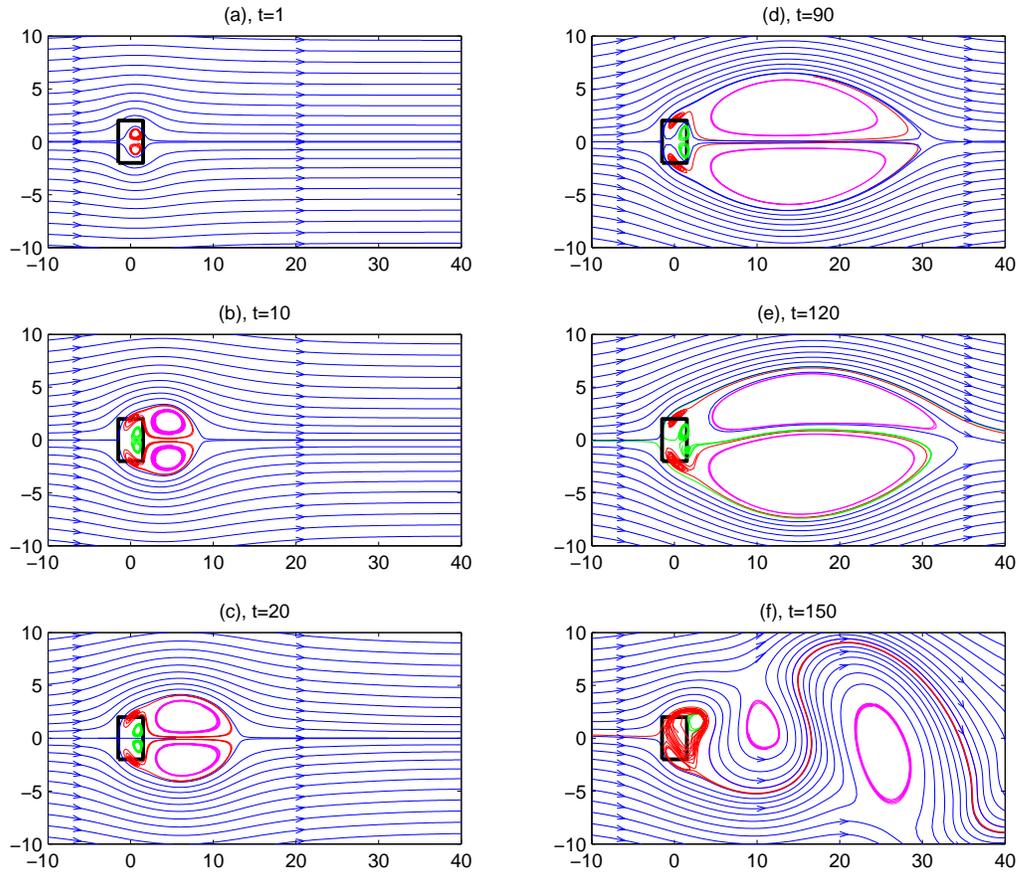}}
    \caption{ \label{Fig:2Dnonstationary} Flow streamlines
    in a 2D nonsteady free flow: $t=1(a)$,
    10($b$), 20($c$), 90($d$), 120($e$), 150($f$);
    ${\bf u(r)}|_{t=0}=(1.5,0)$,  $N\!=\!30$, $Re\!=\!100$.}
\end{figure}

\begin{figure}
    \centerline{\includegraphics[width=9.8cm, angle=0,
    clip=yes]{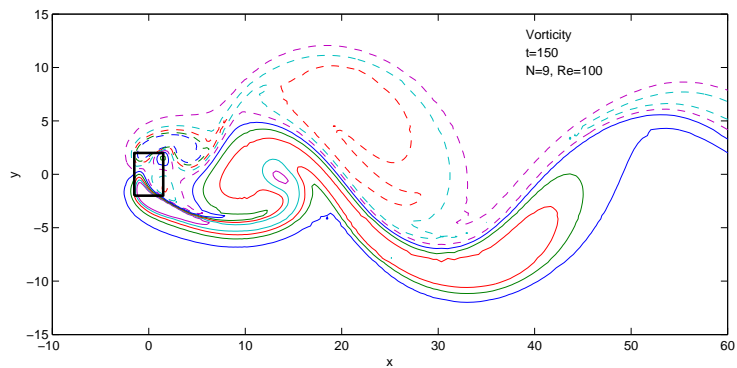}}
    \caption{ \label{Fig:2DVorticity} Vorticity field
    corresponding to Fig.~\ref{Fig:2Dnonstationary}($f$).
    Dashed lines are plotted for negative, and solid lines are plotted
    for positive vorticity.}
\end{figure}

Nevertheless, one can consider \textit{mathematically} what
happens in a 2D free (no side walls) MHD flow where the flow rate
is kept constant. This mathematical problem has been addressed by
\cite{Cuevas:Smolentsev:Abdou:2006}.

The cited 2D MHD flow had no walls, and, hence, no sinks for
kinetic energy except an internal viscosity which is negligible
when $Re$ is high. In such a flow, any vortex pattern cannot be
stabilized, and one would not observe any stationary
recirculation. Instead one obtains the nonsteady vortex
generation, hence, it is worth to discuss a developing flow, i.e.
the flow starting from the constant velocity field ${\bf
u(r)}|_{t=0}=(u_x,0)$. We have performed a few runs of 2D
simulation in order to reproduce results of
\cite{Cuevas:Smolentsev:Abdou:2006} and find out whether the
six-fold vortex pattern shown in Fig.~\ref{Fig:Introductory}($b$)
is of general matter. These results are presented in
Fig.~\ref{Fig:2Dnonstationary}.

One observes, at the initial times, subsequently symmetric
magnetic vortices (Fig.~\ref{Fig:2Dnonstationary}($a$)), and then
a six-vortex pattern (Fig.~\ref{Fig:2Dnonstationary}($b,c$)). This
pattern quickly grows in size due to inertia,
(Fig.~\ref{Fig:2Dnonstationary})($b,c,d$), in such a way that
attached vortices  gradually swell and reach a dimension much
larger than that of magnetic and connecting vortices. When
attached vortices exceed their critical size, they become unstable
and lose symmetry, Fig.~\ref{Fig:2Dnonstationary}($e$). This gives
rise to the Karman vortex street,
Fig.~\ref{Fig:2Dnonstationary}($f$), illustrated also by a
vorticity contour plot in Fig.~\ref{Fig:2DVorticity}. The Karman
vortex street has been reported already by
\cite{Cuevas:Smolentsev:Abdou:2006}, while the preceding temporal
evolution, Fig.~\ref{Fig:2Dnonstationary}($a-f$), is explained
here for the first time.

It is known that the attached vortices in the flow past a solid
body can be discovered  at any Reynolds number $Re$ at the initial
instance of time before vortex shedding starts. One can see that
the same situation takes place in an initial flow past the
magnetic obstacle. The difference is that instead of just attached
vortices one can observe a six-fold vortex pattern developing from
the recirculation under the magnetic gap.

\section{Conclusions}

We have reported the results of a 3D numerical study about a
stationary liquid metal flow in a rectangular duct under the
influence of an external magnetic field. The interaction parameter
$N$, Reynolds number $Re$ as well as magnetic field configuration
have been systematically varied. Whenever it has been possible,
the numerical results have been quantitatively compared with
experimental ones.

First, an analytical, physically consistent and simple model for
the external magnetic field has been derived. Parameters of the
model are geometric dimensions of a region occupied by external
magnets. This model has been successfully verified by comparing
with experimentally measured data and then used through the paper
by varying the constrainment factor, $\kappa=M_y/L_y$, the ratio
between spanwise dimension of the magnet, $2M_y$, and width of the
duct, $2L_y$.

One can classify the following three typical flow structures into
which a stationary MHD flow is organized, depending on the
interaction parameter $N$ as well as spanwise magnetic field
heterogeneity.

The first structure, attributed to low degree of field
heterogeneity, is characterized by the significant electromotive
force which opposes the electrostatic field. The characteristic
pattern for this case is the Hartmann flow. If magnetic field is
uniform, this regime takes place at all $N$, otherwise it is
realized at such $N$ when the field heterogeneity is not strong
manifested.

The second stationary structure is perfectly observed for the
fringing magnetic field, i.e. when the intensity of transverse
magnetic field is varied slowly  in the spanwise and strongly in
the streamwise direction. The flow pattern is given by the
$M$-shaped streamwise velocity profile without recirculation
inside the magnetic gap. Here, the electromotive force and the
electrostatic field can either be opposed or be in the same
direction, but the direction of the electromotive force is always
in the direction of the electric current. The loops of the
electric current are located mainly in the horizontal plane.

The decisive condition for the appearance of the third flow
structure is a strong spanwise variation of the magnetic field
which induces recirculation inside the magnetic obstacle. The
recirculation starts when the reverse electrostatic field prevails
a critical value. Here, the electromotive force is opposed to the
electrostatic field and the direction of the electric current. For
this recirculation regime, the intensities of the reverse flow,
obtained in 3D numerically and by physical experiments have been
compared, and a good agreement has been observed.

The existence regions of stable stationary flow patterns, that is
the dependence of the critical interaction parameter, $N_{c,m}$,
to induce recirculation on the constrainment factor, $\kappa$, has
been calculated and discussed. It has been made clear that no
recirculation is possible for perfectly spanwise uniform external
magnetic field. Moreover, the MHD flows for various $\kappa$ have
been shown to be similar provided they are of the same ratio
$N/N_{c,m}$.

Finally, 3D features of the flow under consideration have been
discussed and it has been demonstrated that the magnetic vortices
are stable in their disposition. This is contrary to a $2D$
numerical study where the stationary recirculation is possible
only in a creeping flow while at higher flow rate the
recirculation develops a vortex shedding. Nevertheless, one can
see all these vortex patterns in a 2D nonsteady flow at initial
times.

We do not want to close this work without some hypotheses on the
nature of the transitions that we have found in this system,
namely, (1) the transition between streamlining flow and the flow
with magnetic vortices when $N_{c,m}$ and $\kappa$ are varied, and
(2) the transition from the two-vortex pattern (only magnetic
vortices) to the six-vortex pattern (magnetic, connecting and
attached vortices) when the Reynolds number is varied. The first
transition has a high probability to be a topological change of
the same solution, which is stable in the whole space of initial
conditions. Increase of the strength as well as the increase or
decrease of the spanwise inhomogeneity of the magnetic field seem
us to be topological changes of the force field resulting in a
topological change of the stationary stable solution. This would
be consistent with the fact, that we never found the different
flow patterns coexisting for the same parameter pair
($N_{c,m}$,$\kappa$). The second transition mentioned is to our
opinion analogous to the appearance of attached vortices behind a
solid obstacle as known from usual hydrodynamics. The magnetic
vortices act as an obstacle for the flow. For increasing Reynolds
number a shear instability arises resulting into the formation of
attached vortices, which for consistence of the flow have to be
accompanied by the connecting vortices. These impressions on the
questions of stability, that we got from our research on the
considered system, nevertheless, remain to be proved rigorously.

%We believe that the recirculation found and discussed in the
%present paper plays the same role for the flow around a magnetic
%obstacle as attached vortices play in ordinary hydrodynamics for
%the flow around a solid body.

\begin{acknowledgments}
 The authors express their gratitude to the Deutsche
Forschungsgemeinschaft for financial support in the frame of the
"Research Group Magnetofluiddynamics" at the Ilmenau University of
Technology under grant ZI 667. The simulations were carried out on
a JUMP supercomputer, access to which was provided by the John von
Neumann Institute (NIC) at the Forschungszentrum J\"{u}lich. We
are grateful for many fruitful discussions with Andre Thess.
Special thanks go to our experimental colleague  Oleg Andreev, for
an always close exchange of thoughts and for providing the
experimental data to compare with our numerical results.
\end{acknowledgments}

\appendix

\section{Indefinite integrals}\label{ap:intergrals}

We take the following notations:  ${\bf r}=(x,y,z)$, $\;{\bf
r'}=(x',y',z')$, $\;\Delta r=|r-r'|=\left[\Delta x^2+\Delta
y^2+\Delta z^2\right]^{1/2}$ and $\Delta x=x-x'$,$\;\Delta
y=y-y'$, $\; \Delta z=z-z'$. Then, the indefinite integration over
$z'$ gives

\begin{eqnarray} \label{app:eq:fz}
    \Phi^{(z)}(x,y,z,x',y',z')=\int \frac{dz'}{|\bf{r-r'}|}=-\mbox{arctanh}\left[\frac{\Delta z}{\Delta
        r}\right],
\end{eqnarray} the indefinite integration over $z',y'$ gives
\begin{eqnarray} \label{app:eq:fzy}
    \Phi^{(z,y)}(x,y,z,x',y',z') &=& \int \frac{dy' dz'}{|\bf{r-r'}|}  =
    - \Delta x\,\mbox{arctan}\left[\frac{\Delta y \Delta z}{\Delta x
    \Delta r}\right] \nonumber \\
        &+& \Delta y\,\mbox{arctanh}\left[\frac{\Delta z}{\Delta
        r}\right]
        + \Delta z\,\mbox{arctanh}\left[\frac{\Delta y}{\Delta r}\right],
\end{eqnarray} and finally the indefinite integration over $z',y',x'$ gives

\begin{eqnarray} \label{app:eq:fzyx}
    \Phi^{(z,y,x)}(x,y,z,x',y',z') &=& \int \frac{dx' dy' dz'}{|\bf{r-r'}|}  \nonumber \\
    &=&
        \frac{1}{2}\Delta x^2\arctan\left[\frac{\Delta y \Delta z}{\Delta x \Delta
        r}\right]
        -\Delta y\Delta z\, \mbox{arctanh}\left[\frac{\Delta x}{\Delta
        r}\right] \nonumber \\
    &+&
        \frac{1}{2}\Delta y^2\arctan\left[\frac{\Delta x \Delta z}{\Delta y \Delta
        r}\right]
        -\Delta x\Delta z\, \mbox{arctanh}\left[\frac{\Delta y}{\Delta
        r}\right] \nonumber \\
    &+&
        \frac{1}{2}\Delta z^2\arctan\left[\frac{\Delta x \Delta y}{\Delta z \Delta
        r}\right]
        -\Delta x\Delta y\, \mbox{arctanh}\left[\frac{\Delta z}{\Delta
        r}\right].
\end{eqnarray}

To confirm, one can use a commercial program, e.g. Mathematica, to
analytically differentiate (\ref{app:eq:fzyx}) back and get
(\ref{app:eq:fz}) finally. The original way to calculate
(\ref{app:eq:fz})-(\ref{app:eq:fzyx}) is quite cumbersome, and is
not provided here.

%For instance, the integration over a finite 3D rectangular block
%$\Omega=\{x_1\!\leq\!x'\!\leq\!x_2,\; y_1\!\le\!y'\!\le\!y_2,\;
%z_1\!\le\!z'\!\le\!z_2\!\}$ takes the following form
%\begin{eqnarray}
%\int_\Omega\frac{d
%    \bf{r'}}{|\bf{r-r'}|}=\left[\left[\left[\Phi^{(z,y,x)}(x,y,z,x',y',z')
%    \right]_{x'=x_1}^{x'=x_2}
%    \right]_{y'=y_1}^{y'=y_2}
%    \right]_{z'=z_1}^{z'=z_2}.
%\end{eqnarray}

\bibliographystyle{jfm}
% Note the spaces between the initials
\bibliography{./../../Bibtex/mhd}

\end{document}